\documentclass[a4paper]{jpconf}

\usepackage{graphicx}
\usepackage{subfigure}
\usepackage{amsmath,color,bm,iopams,afterpage}
\def\gappeq{\mathrel{ \rlap{\raise.5ex\hbox{$>$}}
                      {\lower.5ex\hbox{$\sim$}} } }
\def\lappeq{\mathrel{ \rlap{\raise.5ex\hbox{$<$}}
                      {\lower.5ex\hbox{$\sim$}} } }

\usepackage{color}

\newcommand{\del}[1]{\textcolor{red}{}}

\providecommand{\Rey}{\mbox{\ensuremath{\mathcal{R}}}}
\newcommand{\hidden}[1]{}

\begin{document}

\title{Classical-like wakes past elliptical obstacles in atomic Bose-Einstein condensates}

\author{G. W. Stagg, A. J. Allen, C. F. Barenghi and N. G. Parker} 
\address{Joint Quantum Centre Durham--Newcastle, School of Mathematics and Statistics, Newcastle University, Newcastle upon Tyne, NE1 7RU, United Kingdom}
\ead{nick.parker@newcastle.ac.uk}

\begin{abstract}
We reinvestigate numerically the classic problem of two-dimensional superfluid flow past an obstacle.  Taking the obstacle to be elongated (perpendicular to the flow), rather than the usual circular form, is shown to promote the nucleation of quantized vortices, enhance their subsequent interactions, and lead to wakes which bear striking similarity to their classical (viscous) counterparts.  Then, focussing on the recent experiment of Kwon {\it et al.} (arXiv:1403.4658) in a trapped condensate, we show that an elliptical obstacle leads to a cleaner and more efficient means to generate two-dimensional quantum turbulence.
\end{abstract}

\section{Introduction}

A prototypical example in classical (viscous) fluid dynamics is the incompressible flow past a cylinder, depicted in Fig. 1.  The nature of the flow is determined by the Reynolds number $\Rey = vd/\nu$, where $v$ is the flow velocity away from the obstacle, $d$ is the obstacle diameter,
and $\nu$ is the fluid's
kinematic viscosity.  For $\Rey \lesssim 5$ the flow is laminar around the obstacle.  For $5 \lesssim \Rey\lesssim50$ a steady symmetric wake forms downstream of the obstacle.  For $10^2\lesssim\Rey\lesssim10^5$ the wake 
becomes asymmetric and time dependent, forming the famous 
B\'enard--von K\'arm\'an vortex street structure.  At even higher $\Rey$,
the wake becomes turbulent. 

\begin{figure}[h]
\centering
  \includegraphics[height=2.25cm,angle=180]{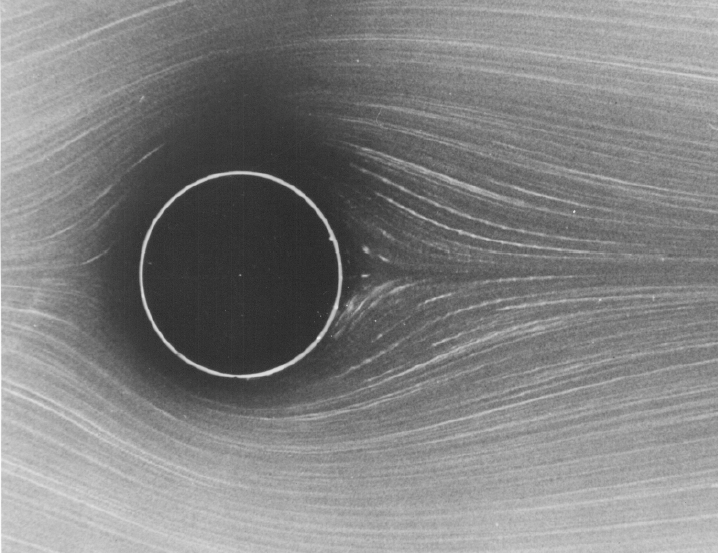}
  \includegraphics[width=0.24\textwidth,angle=180]{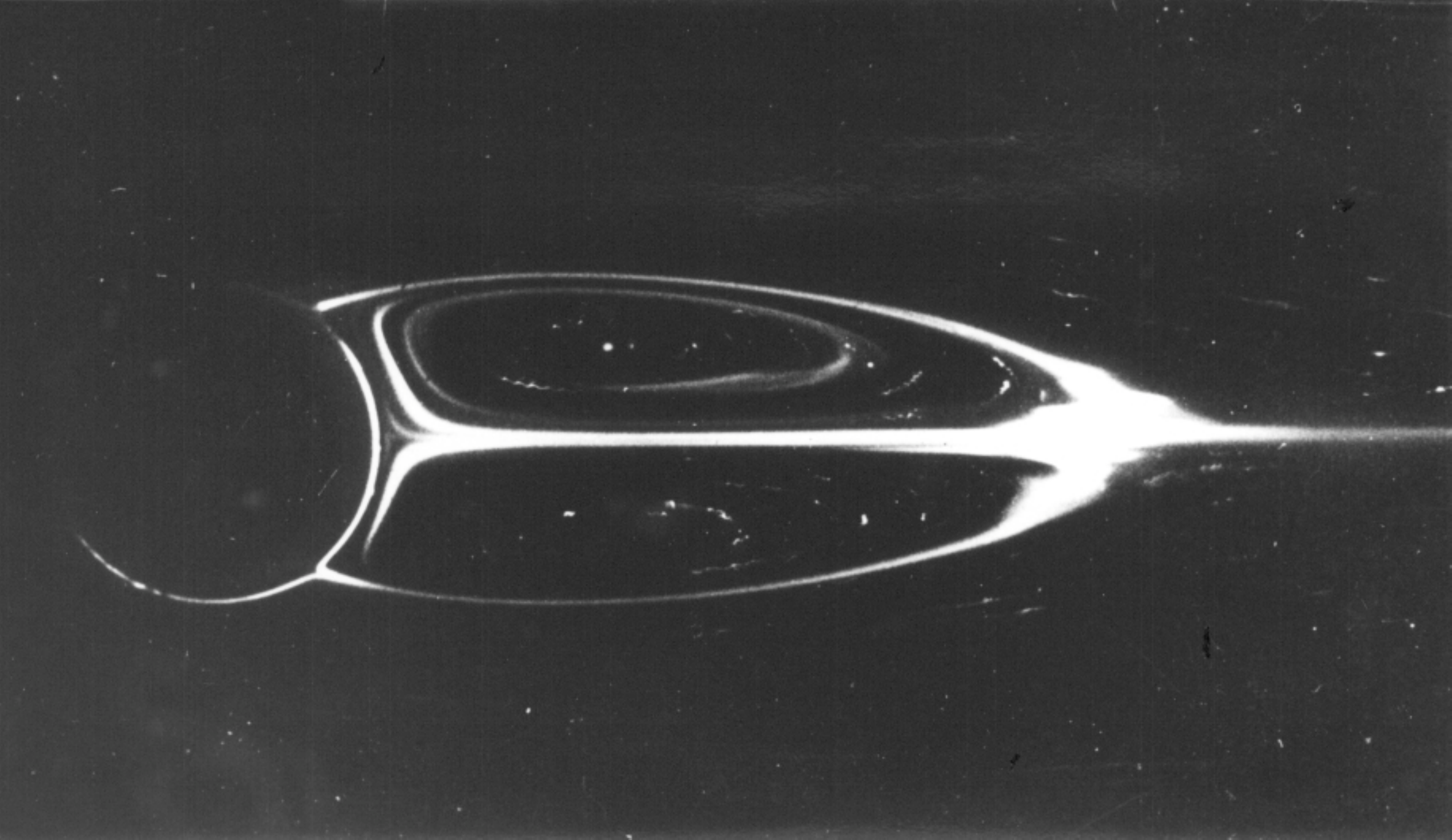}
  \includegraphics[width=0.24\textwidth,angle=180]{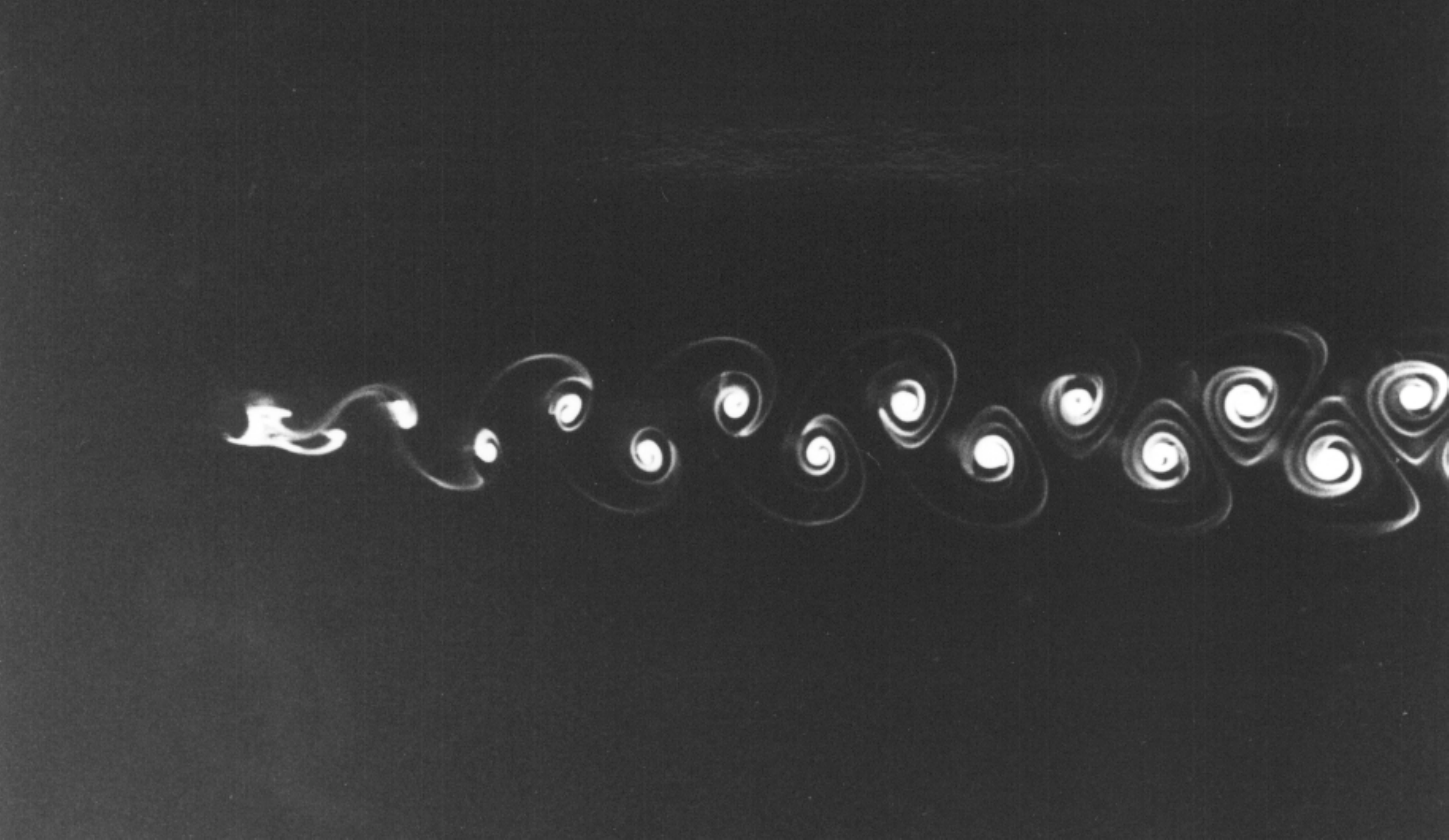}
    \includegraphics[width=0.29\textwidth,angle=180]{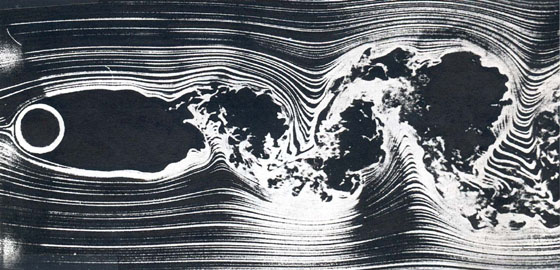}
  \caption{Classical viscous flow past a cylinder. From left to right: laminar flow ($\Rey=3.64$) \cite{taneda41}; steady symmetric wake behind the cylinder ($\Rey=41$) \cite{taneda41}; time-dependent B\'enard--von K\'arm\'an vortex street ($\Rey = 112$) \cite{taneda112}; and chaotic downstream wake ($\Rey > 10^5$) \cite{nagib}.} 
  \label{fig:taneda-imgs}
\end{figure}

In a superfluid the situation is fundamentally different \cite{annett}.  There is no viscosity so a Reynold's number cannot be defined. Landau's criterion states that dissipation through elementary excitations kicks in only above a critical velocity; for the dilute Bose gas relevant here this critical velocity is approximately the speed of sound.  Meanwhile, where eddies do exist in the fluid, they are constrained by quantum mechanics to possess quantized vorticity and circulation.  In their pioneering simulations of the 2D nonlinear Schr\"odinger equation (NLSE), Frisch {\it et al.} \cite{frisch92} showed that superflow past a circular obstacle becomes dissipated through the nucleation of quantized vortices at the object, which then get carried downstream.  

Atomic Bose-Einstein condensates provide a clean and highly controllable vehicle for studying superfluid phenomena, from persistent currents \cite{persistent} and Josephson junctions \cite{jo} to quantized vorticity \cite{vortices} and quantum turbulence \cite{Henn,Neely, kwon_moon_14,qt}.  Various experiments have probed the response of the condensate to a moving obstacle \cite{Neely,kwon_moon_14,Raman,Onofrio,Inouye}.  The obstacle in these cases takes the form of a repulsive Gaussian-shaped potential created by a blue-detuned laser beam.  Of these experiments, \cite{Neely,kwon_moon_14} employed a highly flattened condensate such that the dynamics were effectively two-dimensional and provided a close analog of the cylindrical problem.  Moreover, they employed this scenario to generate two-dimensional quantum turbulence, a dynamical state of the condensate characterized by a statistical distribution of excitations over a large range of lengthscales.

In the mean-field picture, such a 2D condensate can be parameterized by a macroscopic wavefunction $\psi(x,y,t)$ whose behaviour follows the 2D Gross-Pitaevskii equation \cite{stringari},  
\begin{equation}
i \hbar \frac{\partial\psi}{\partial t} = \left(-\frac{\hbar^2}{2m}\nabla^2 + V(x,y,t)+ g|\psi|^2 - \mu \right) \psi.
\label{eq:gpe1}
\end{equation} 
Here $g=4 \pi \hbar^2 a_{\rm s}/\sqrt{2 \pi l_z} m$ is a nonlinear coefficient arising from the contact atomic interactions, with $a_{\rm s}$ being the atomic {\it s}-wave scattering length and $m$ the atomic mass, $\mu$ is the chemical potential of the condensate, and $V(x,y,t)$ specifies the potential exerted on the condensate.

\begin{figure}[h]
\centering
	~~~(a) \hspace{6.5cm} (b) ~~~~~~~~~~~~~~~~~~~~~~~~~~~~~~~~~~~~~~~~~~~~~~~~~~~~~~~
	\\
	\includegraphics[width=0.425\linewidth]{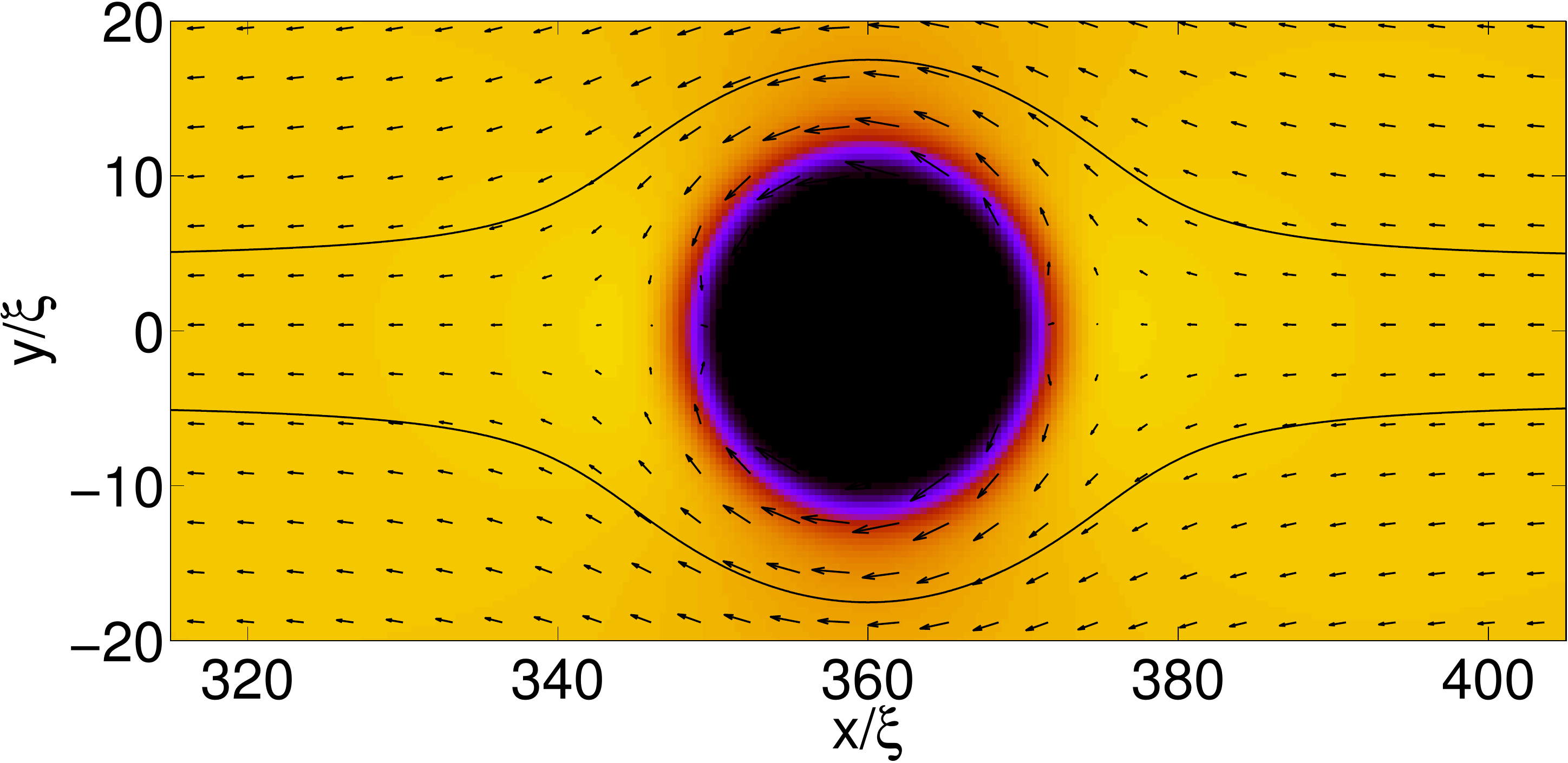} \hspace{0.2cm}
	\includegraphics[width=0.425\linewidth]{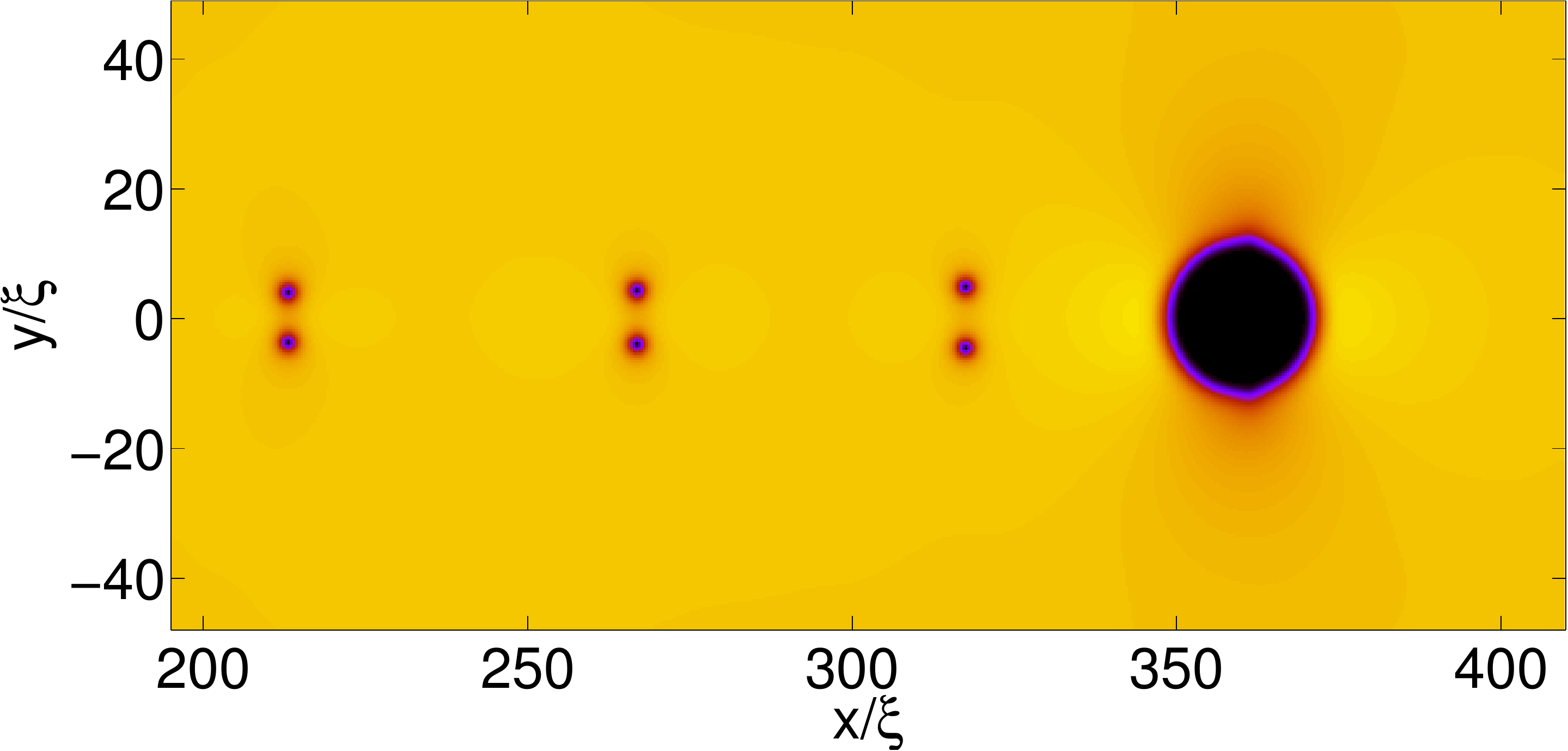}\\
	~~~(c) \hspace{6.5cm} (d)~~~~~~~~~~~~~~~~~~~~~~~~~~~~~~~~~~~~~~~~~~~~~~~~~~~~~~~
	\\
	\includegraphics[width=0.425\linewidth]{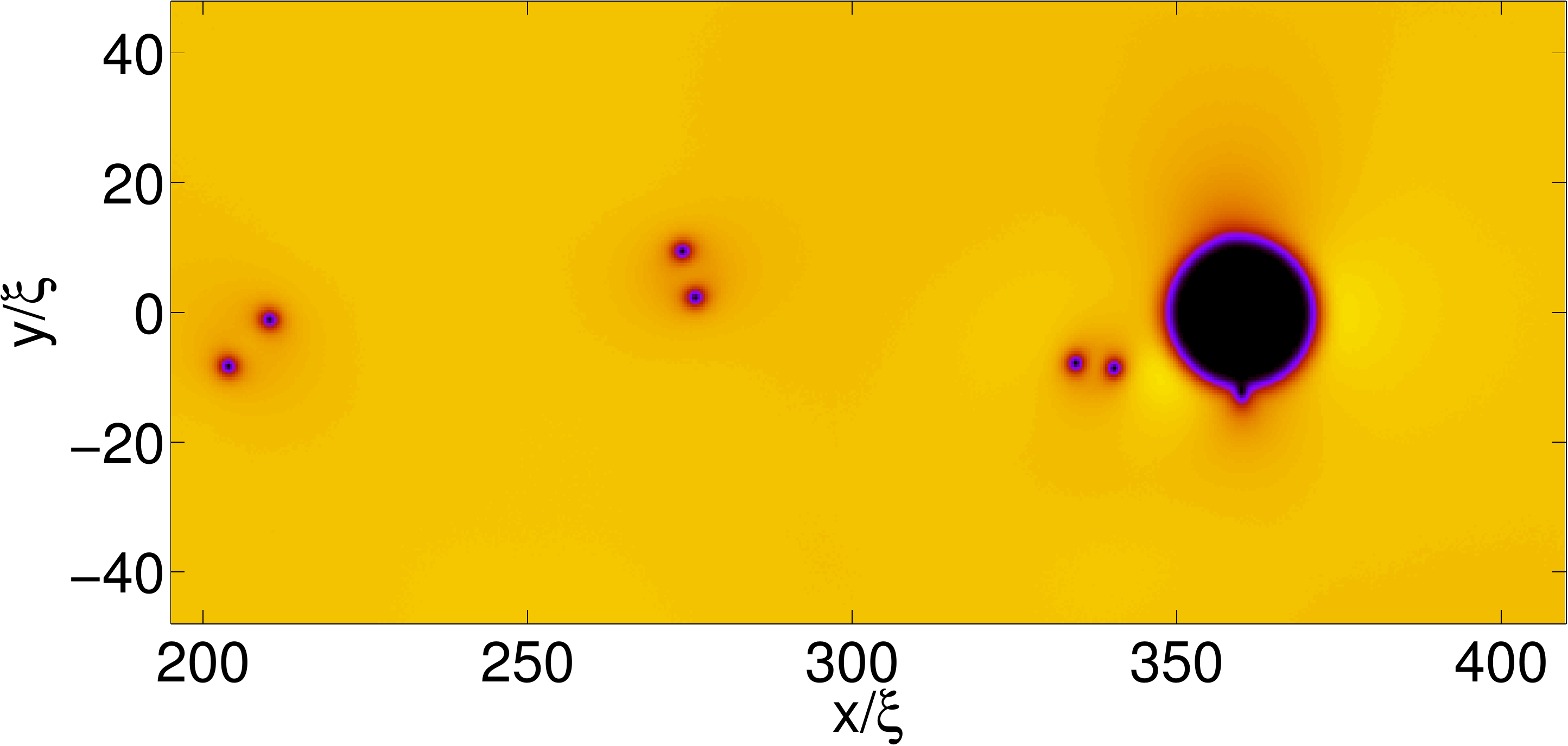} \hspace{0.2cm}
	\includegraphics[width=0.425\linewidth]{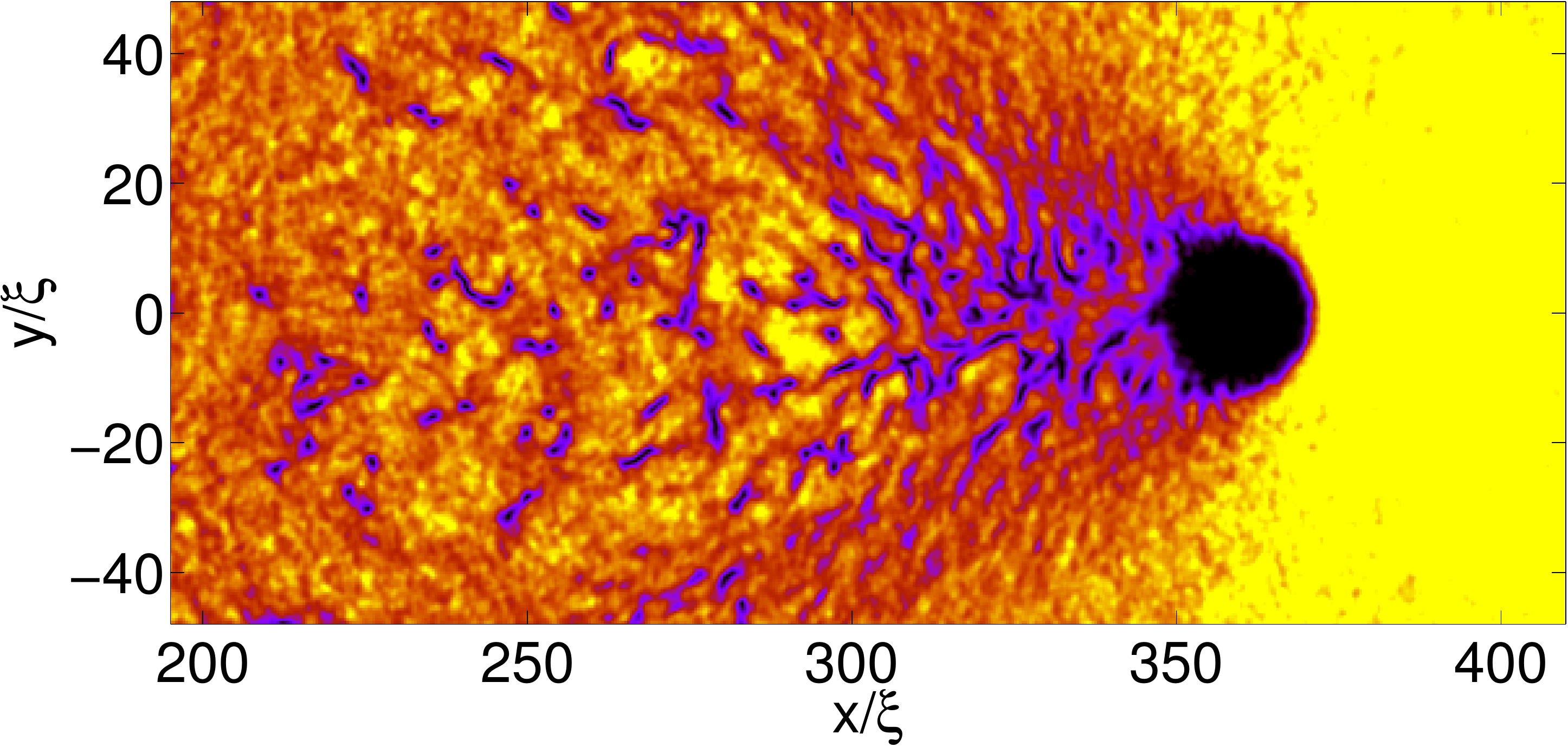}
	\caption{\label{fig:denstraj} Condensate density during flow (in the positive $x$-direction) past a circular obstacle ($d = 5\xi$) at various flow speeds. Yellow/black represent high/low density.  For reference the critical velocity here is $v_c \approx 0.36c$.  (a) Laminar flow at a sub-critical flow speed ($v=0.3c$).  The fluid velocity vector field and two streamlines illustrate the flow pattern. (b) Nucleation of vortex-antivortex pairs ($v=0.365c$).  (c) Nucleation of like-signed vortex pairs ($v=0.37c$). (d)  Chaotic vortex nucleation and generation of strong sound waves, forming a turbulent wake ($v=0.9c$).   Length and speed are expressed in terms of the healing length $\xi = \hbar/\sqrt{m n_0 g}$ and speed of sound $c=\sqrt{n_0 g/m}$.}
\end{figure}

Whereas Frisch {\it et al.} \cite{frisch92} considered a ``hard'' cylinder, it is more natural in the context of atomic condensates to employ a ``soft'' Gaussian potential of the form $V(x,y)=V_0 e^{-(x^2+y^2)/d^2 }$, where $d$ and $V_0$ parameterize the width and amplitude of the potential.  In practice this changes the quantitative, but not the qualitative, behaviour.  Sasaki {\it et al.} \cite{saito10} recently provided an extensive picture of 2D superflow past such a Gaussian obstacle.   The flow regimes are depicted in Fig. \ref{fig:denstraj}, based on simulations of the 2D GPE using similar parameters to \cite{saito10}.   At low flow velocity [(a)], the fluid undergoes smooth laminar flow around the obstacle.  The streamlines of this flow are symmetric about $x=0$, as in perfect potential flow.  At a critical flow velocity, the local fluid velocity (which is highest at the poles of the obstacle) exceeds the speed of sound, breaking Landau's criterion.  Vortex pairs of opposite sign are nucleated periodically and drift downstream, forming a collimated wake of vortex pairs which are widely separated from each other [(b)].  For higher velocities, alternating pairs of like-signed vortices are nucleated [(c)].  At higher velocities, vortex nucleation becomes highly irregular [(d)], forming a chaotic downstream distribution of vortices and sound (density) waves.  These quantum fluid flow patterns bear some analogy to the classical flow patterns of Fig. 1, particularly for the laminar flow and chaotic regimes.  The alternating like-signed vortices form a somewhat primitive analog of the Bern\'ard-von K\'arm\'an vortex street, while the vortex-antivortex pairs have no obvious classical analog.  

In this work we discuss the rich variety of quantum wake regimes, often in close analog to the classical counterparts, which can be obtained via the simple modification of the obstacle to an {\it elliptical} shape.  In other words, we employ an elliptical potential of the form $V(x,y)=V_0 e^{-(\varepsilon^2 x^2+y^2)/d^2 }$, where $\varepsilon$ is the ellipticity.  This is readily achieved experimentally by using cylindrical focussing (rather than spherical focussing) of the laser beam which induces the potential.  First, we explore these dynamics in a homogeneous system, which serves to demonstrate the salient behaviour of superfluid flow past an elliptical obstacle, away from boundaries and trap-induced density inhomogeneities which also influence the vortex dynamics.  Then we consider the realistic scenario provided by the recent experiment of Kwon {\it et al.} \cite{kwon_moon_14}, and investigate how the dynamics therein are modified by an elliptical obstacle.

\section{Flow Past an Elliptical Obstacle: Homogeneous Condensate}
In the following we simulate the 2D GPE for an untrapped 2D condensate with uniform density $n_0$,  flowing at speed $v$ along $x$ (modelled via the addition of a Galilean term $ i\hbar v\dfrac{\partial}{\partial x} \psi$ to the right-hand side of the GPE (\ref{eq:gpe1})). A static elliptical obstacle potential, elongated in $y$ ($\varepsilon>1$) and with arbitrary size $d$ (expressed in terms of the healing length $\xi=\hbar/\sqrt{m n_0 g}$), pierces the condensate.  Further technical details are provided elsewhere \cite{stagg_parker_14}.  Note that the qualitative behaviour in insensitive to additional noise.  %All of these results are based on simulations of the 2D GPE, and we have verified that the qualitative dynamics are not affected by the inclusion of additive noise or dissipation (through a phenomenological dissipation term to the GPE \cite{stagg_allen_14}). 

 % Unless stated otherwise, a small amount of noise is added to the initial condition to break symmetry: a random number between $-0.0005$ and $0.0005$ is added to both the real and imaginary parts of the initial wavefunction. 

In the following we describe the behaviour at increasing flow speed $v$.  At sub-critical speeds, the condensate undergoes laminar flow around the obstacle, much like for the circular obstacle [Fig. 2(a)].  However, due to the decreased radius of curvature at the poles of the obstacle, the local fluid velocity is higher there than for the circular obstacle.  As such the critical velocity is reduced with increasing $\varepsilon$, as revealed by GPE simulations (data points in Fig. 3).    Approximating the BEC as an inviscid Euler fluid, the maximum local fluid velocity is $v_{\rm max}=(1+\varepsilon)v$, which will exceed the speed of sound $c$ at a critical flow speed  $v_{c} = c/(1+\varepsilon)$.  Despite its crudeness, this approximation (black dashed line in Fig. 3) predicts the correct scaling with $\varepsilon$.  Inclusion of higher-order effects, e.g. the reduction of density around the obstacle due to the raised fluid speed (Bernoulli's theorem) \cite{win01,stagg_parker_14}, bring the model into more precise quantiative agreement (red solid line in Fig. 3).
\begin{figure}[h]
\centering
\includegraphics[width=0.5\textwidth]{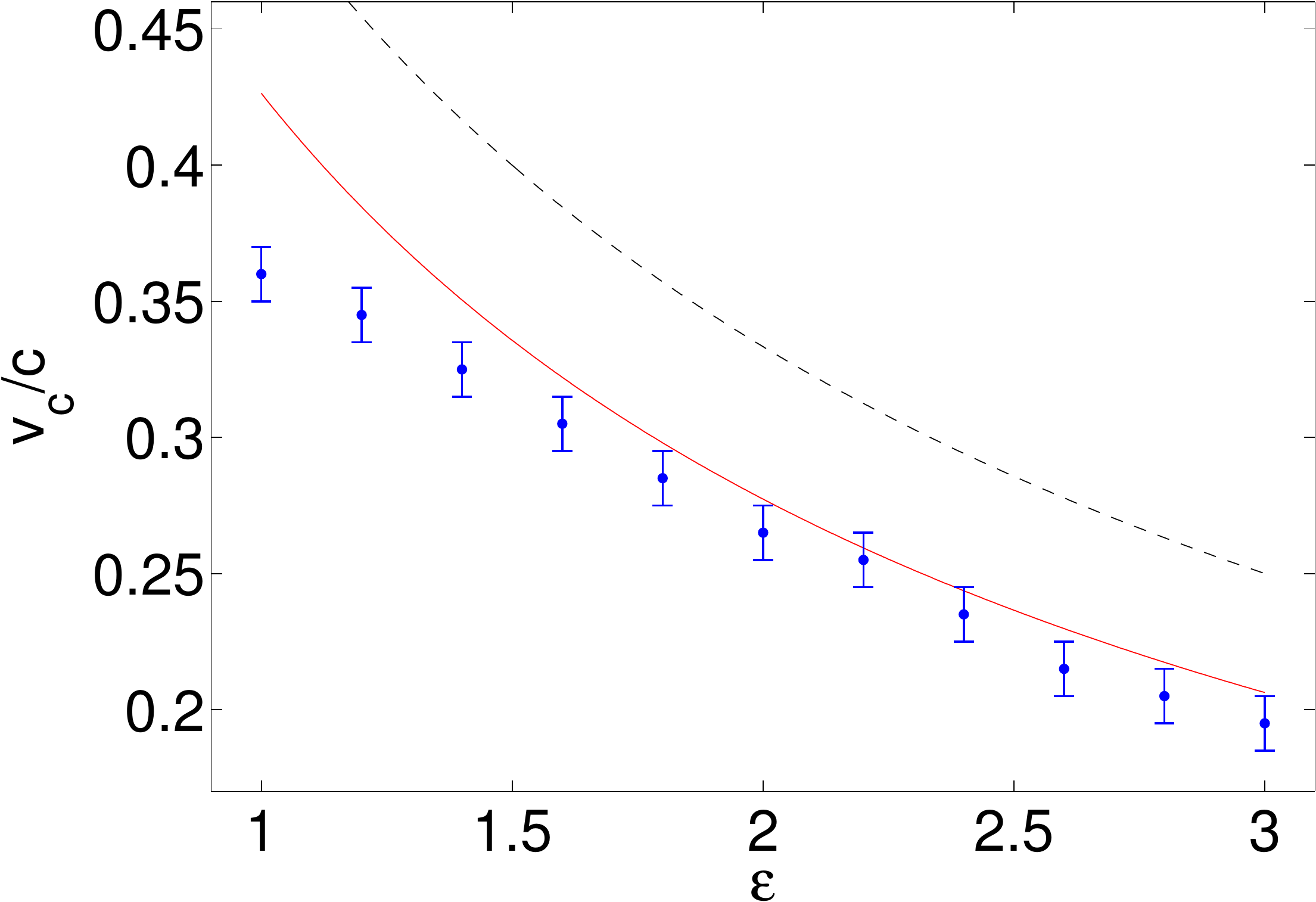} 
\caption{\label{fig:velplots}Critical velocity against obstacle ellipticity $\varepsilon$ ($d=10\xi$) according to GPE  simulations (blue bars), the Euler prediction $v_c=c/(1+\varepsilon)$ (black dashed), and the corrected Euler prediction \cite{win01,stagg_parker_14} (solid red).
}
\end{figure}

For $v>v_c$ the ellipticity also increases the rate at which vortices are nucleated.  Importantly, a large vortex nucleation rate implies that vortices are close together and thereby interact strongly. (For the circular obstacle, while a higher rate of vortex shedding can be achieved by using a faster flow, this also carries the vortices downstream more quickly, and so the vortex interactions are not significantly increased.)  
\begin{figure}[b]
\centering
	(a) \hspace{7.2cm} (b) ~~~~~~~~~~~~~~~~~~~~~~~~~~~~~~~~~~~~~~~~~~~~~~~~~~~~~~~
	\\
	\includegraphics[width=0.45\linewidth]{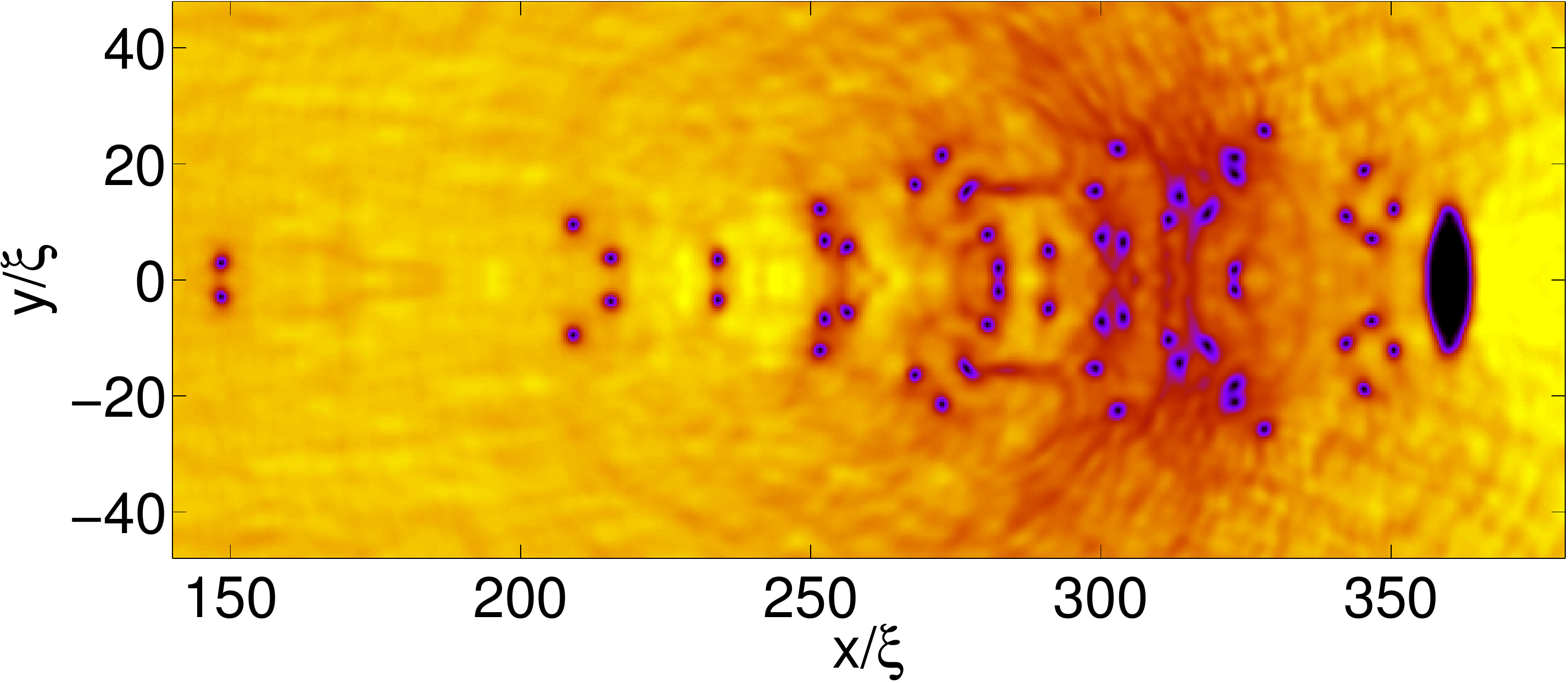} \hspace{0.5cm}
	\includegraphics[width=0.45\linewidth]{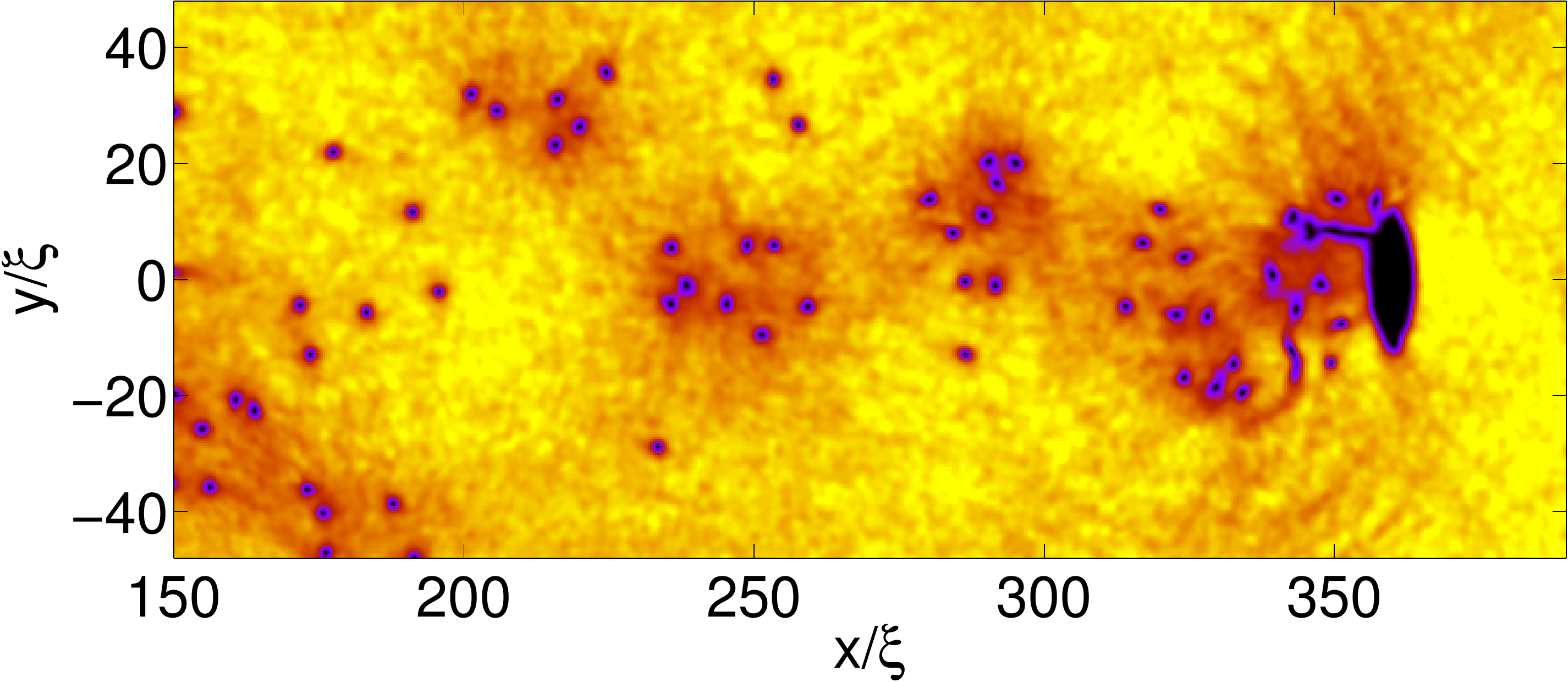}\\
		(c) \hspace{7.2cm} (d) ~~~~~~~~~~~~~~~~~~~~~~~~~~~~~~~~~~~~~~~~~~~~~~~~~~~~~~~
	\\
	\includegraphics[width=0.45\linewidth]{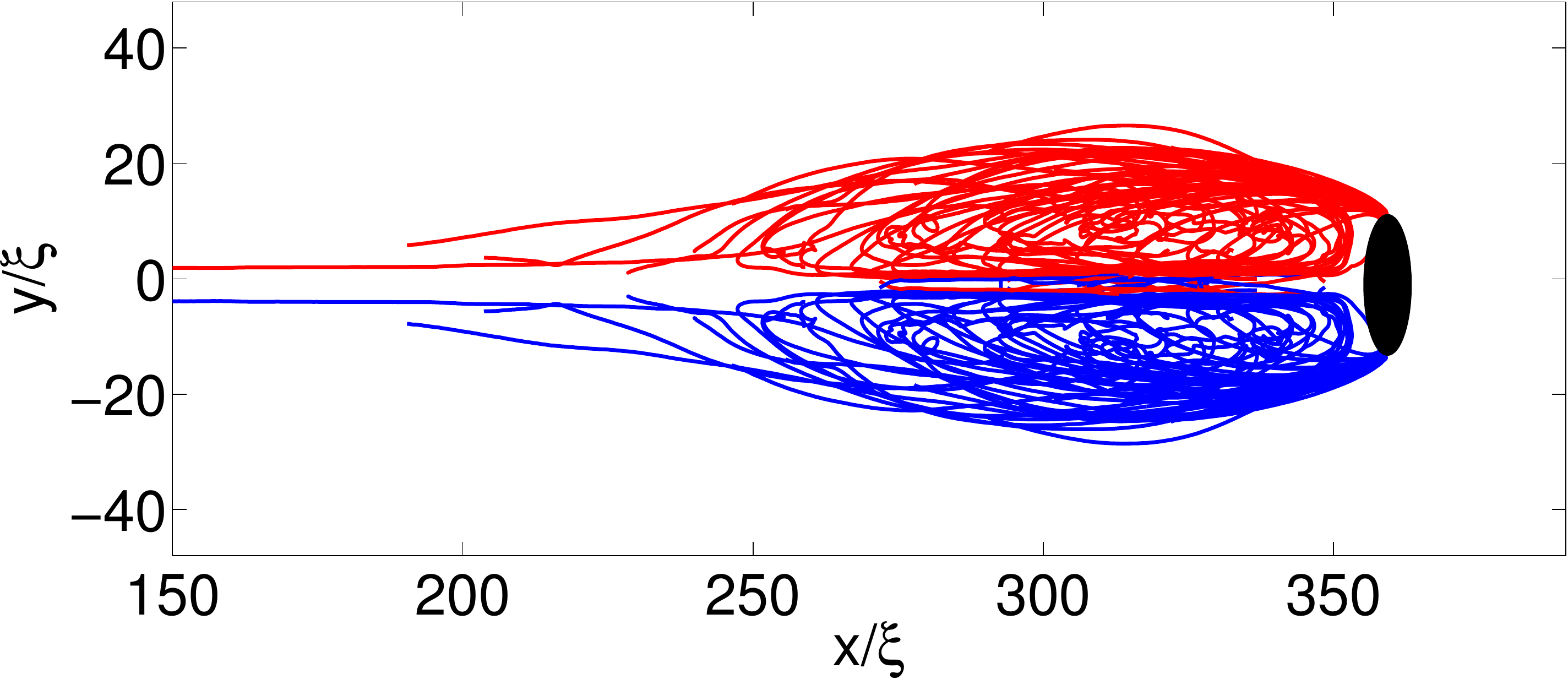} \hspace{0.5cm}
	\includegraphics[width=0.45\linewidth]{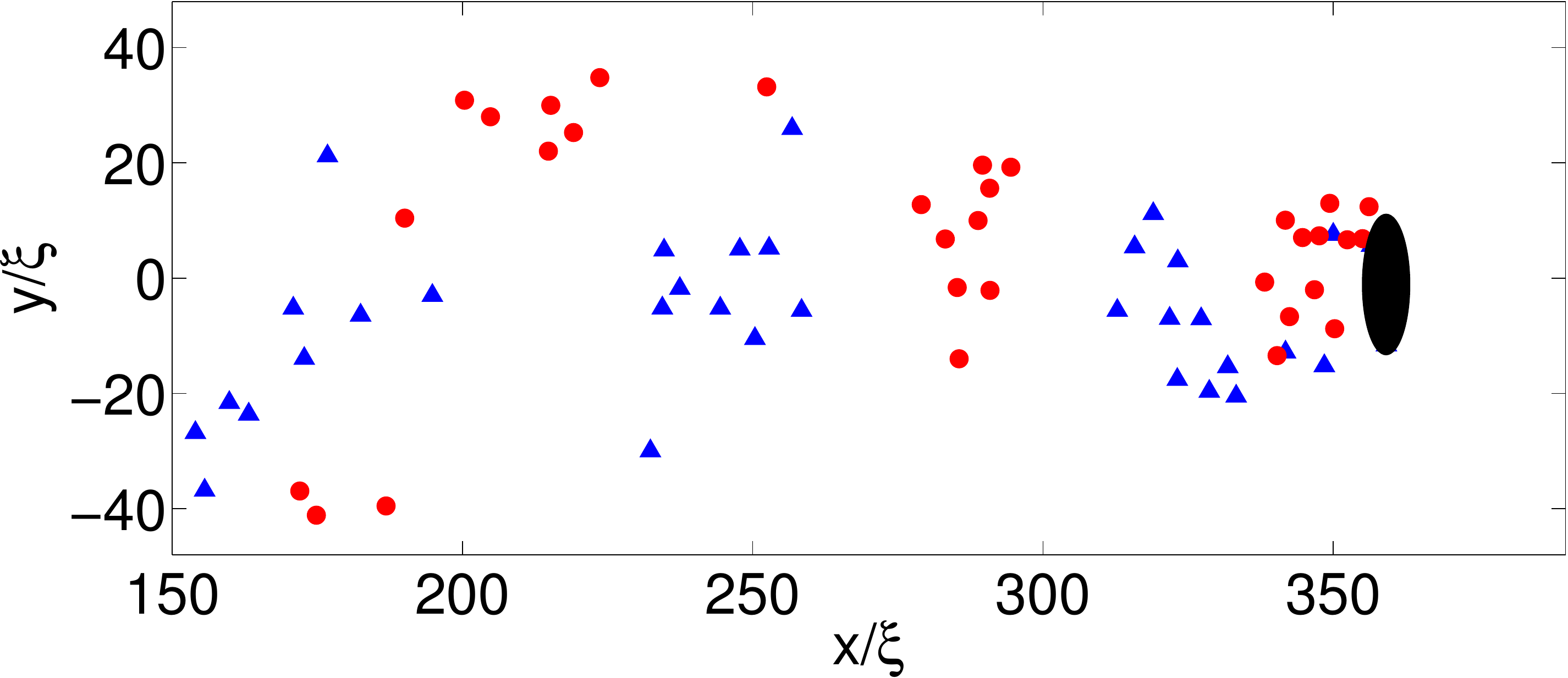}
	\caption{\label{fig:N_vV2} Condensate flow ($v=0.52c$) past an elliptical obstacle ($d=5\xi$, $\varepsilon=3$). (a) Symmetric wake.  (b) Asymmetric wake.  (c) Time-integrated vortex trajectories for the symmetric wake.  (d) Vortex positions for the asymmetric wake shown in (b).    Red and blue represent vortices of opposite circulation.  }
\end{figure}

The rapid nucleation of vortex-antivortex pairs leads to a cluster of vortices in the one half-plane and antivortices in the other half-plane [Fig. 4(a)].  In each cluster, the same-sign vortices tend to corotate around each other, such that the cluster approximates a macroscopic  vortex.  A steady-state is reached whereby the freshly-nucleated vortices entering the clusters is balanced by vortex pairs which escape the cluster downstream.    The close analogy to classical flow becomes apparent in plotting the trajectories of the vortices [Fig. 4(c)]; two symmetric lobes of circulating vortices are revealed, in striking similarity to the symmetric classical wake in Fig. 1.  

While the above flow pattern is symmetric in $y$, it is dynamically unstable to the formation of an asymmetric pattern.  A typical snapshot of this is shown in Fig. 4(b), with Fig. 4(d) indicating the corresponding position and circulation of the vortices.  This pattern is characterised by the nucleation of several like-signed vortices from one pole of the obstacle, followed by nucleation of several vortices of the opposite sign from the opposite pole.  The result is a wake consisting of alternating clusters of like-signed vortices.  The number of vortices in each cluster increases with increasing $v$, $d$ and $\varepsilon$.  This wake is a strongly analogous to the classical B\'enard--von K\'arm\'an vortex street shown in Fig. 1(c).  Indeed, if one imagines course-graining over the clusters, one obtains a flow pattern of the form of the B\'enard--von K\'arm\'an vortex street.  At even higher flow speeds, the vortex nucleation becomes irregular, forming a chaotic wake like that presented for the circular obstacle in Fig. 2(d). 

Overall, we see that the superfluid flow past an elliptical obstacle leads to four flow pattern regimes, each of which is bears striking analogy to the classical viscous cases depicted in Fig. 1.  These findings confirm the intuition that a sufficiently large quanta of vorticity/circulation reproduces classical results.

\section{Flow Past an Elliptical Obstacle: Trapped Condensate}
\afterpage{
\clearpage
\begin{figure}
	\begin{centering}
	\includegraphics[width=0.22\linewidth,clip=true,trim=70 40 0 40]{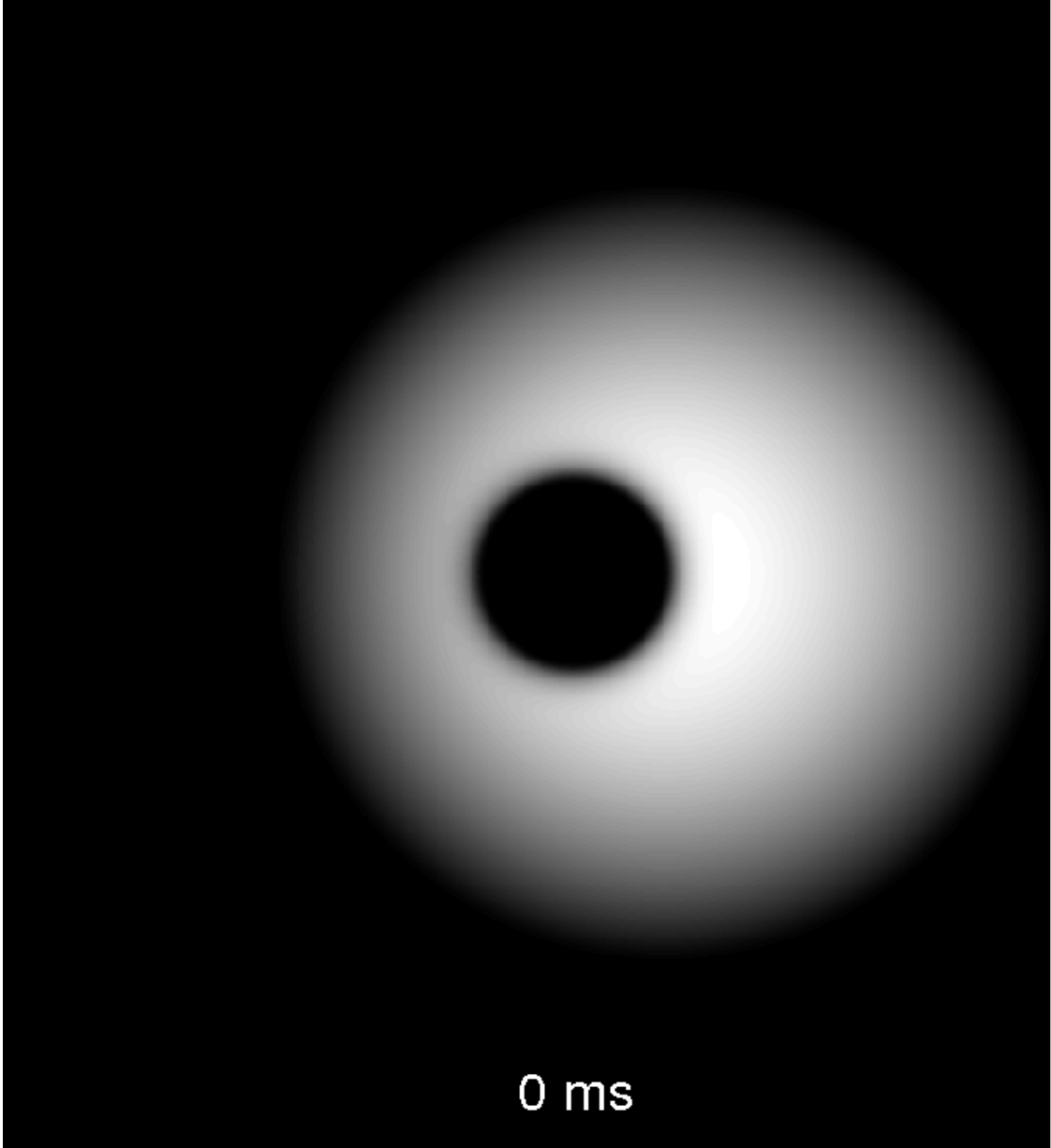}
	\hspace{-0.52in}\color{white}\raisebox{0.5ex}{\textsf{000~ms}}
	\includegraphics[width=0.22\linewidth,clip=true,trim=05 40 65 40]{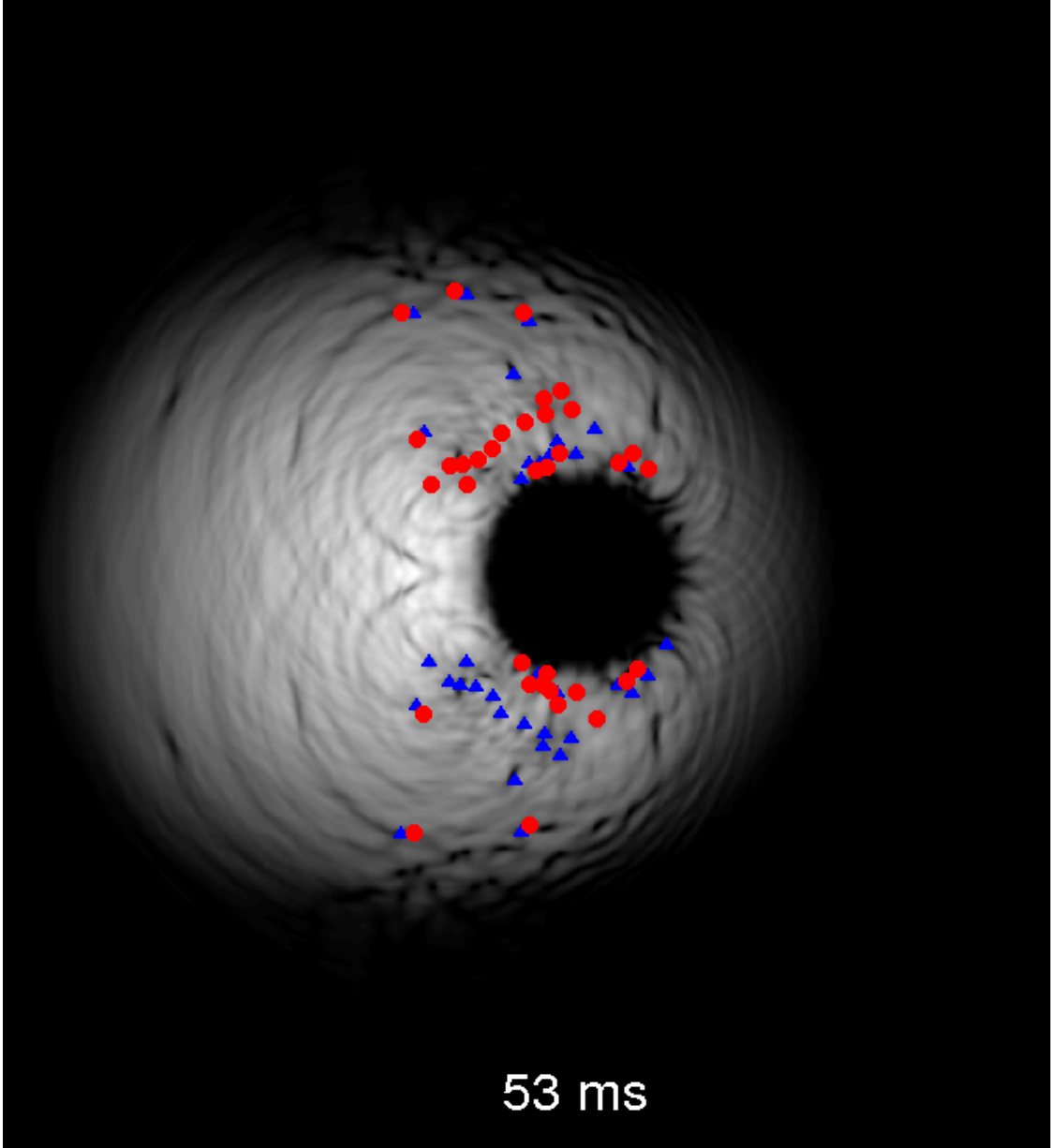}
	\hspace{-0.52in}\color{white}\raisebox{0.5ex}{\textsf{053~ms}}
	\includegraphics[width=0.22\linewidth,clip=true,trim=50 40 20 40]{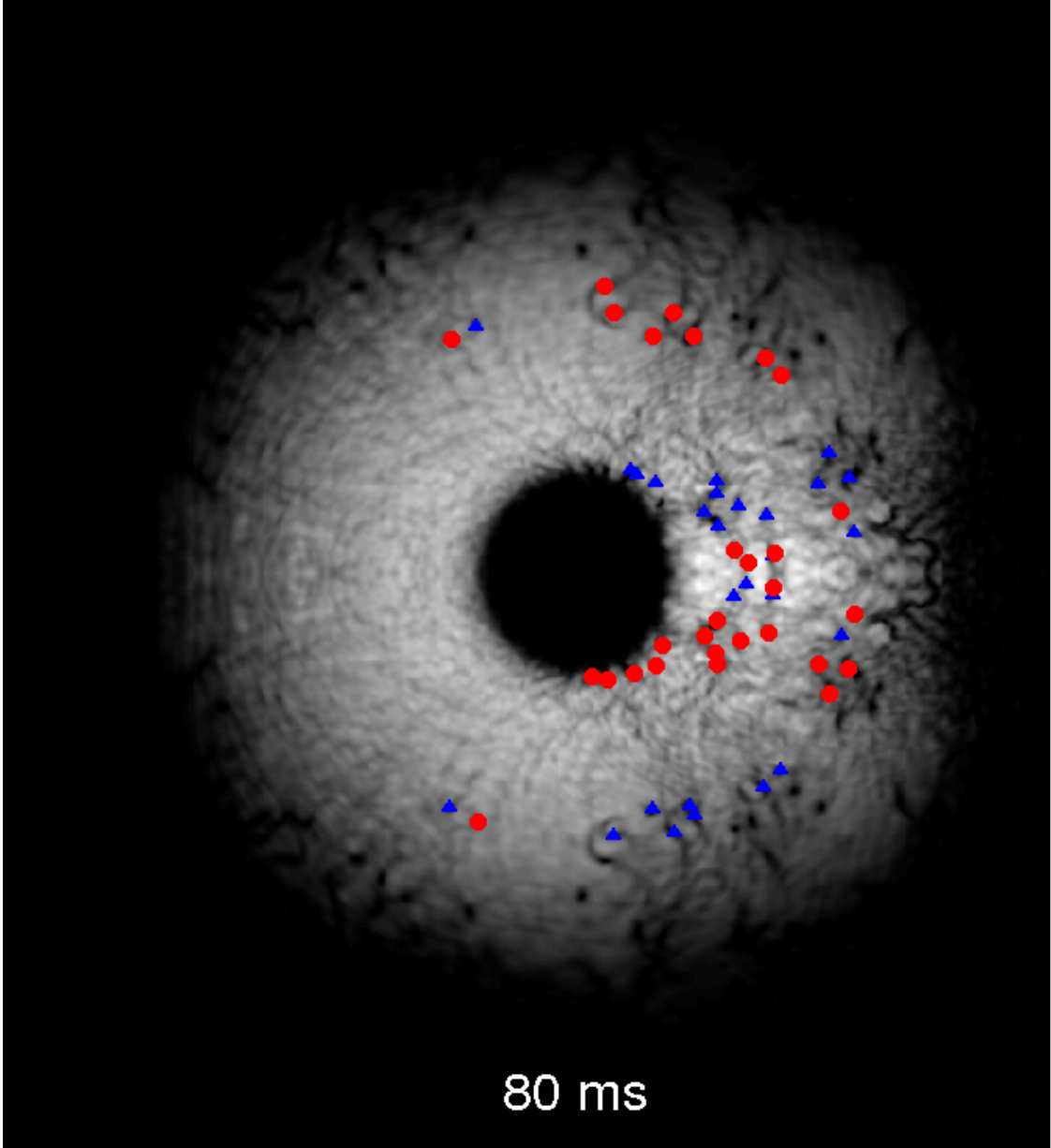}
	\hspace{-0.52in}\color{white}\raisebox{0.5ex}{\textsf{080~ms}}
	\includegraphics[width=0.22\linewidth,clip=true,trim=20 40 50 40]{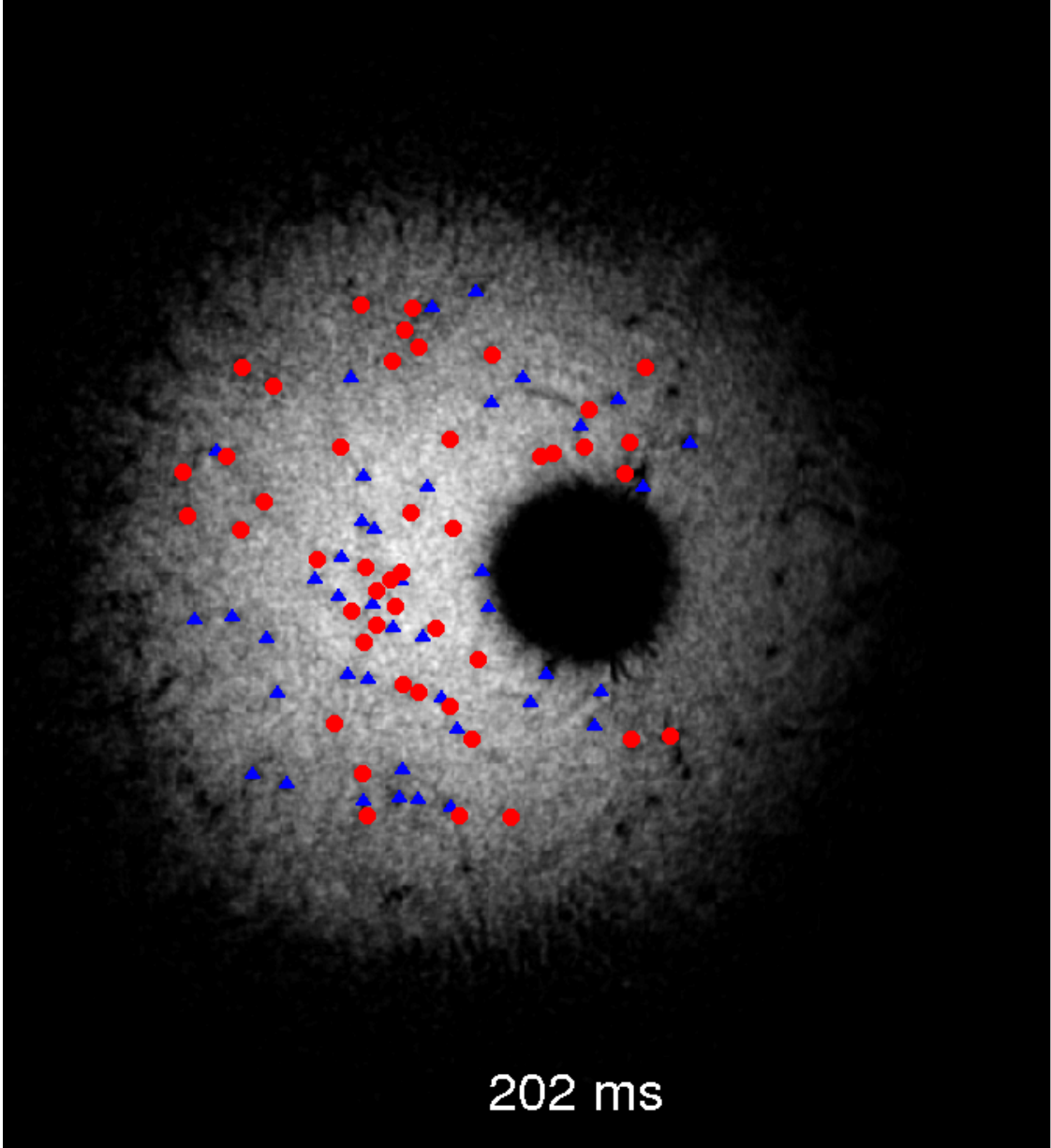}
	\hspace{-0.52in}\color{white}\raisebox{0.5ex}{\textsf{202~ms}}\\
	\includegraphics[width=0.22\linewidth,clip=true,trim=45 40 25 40]{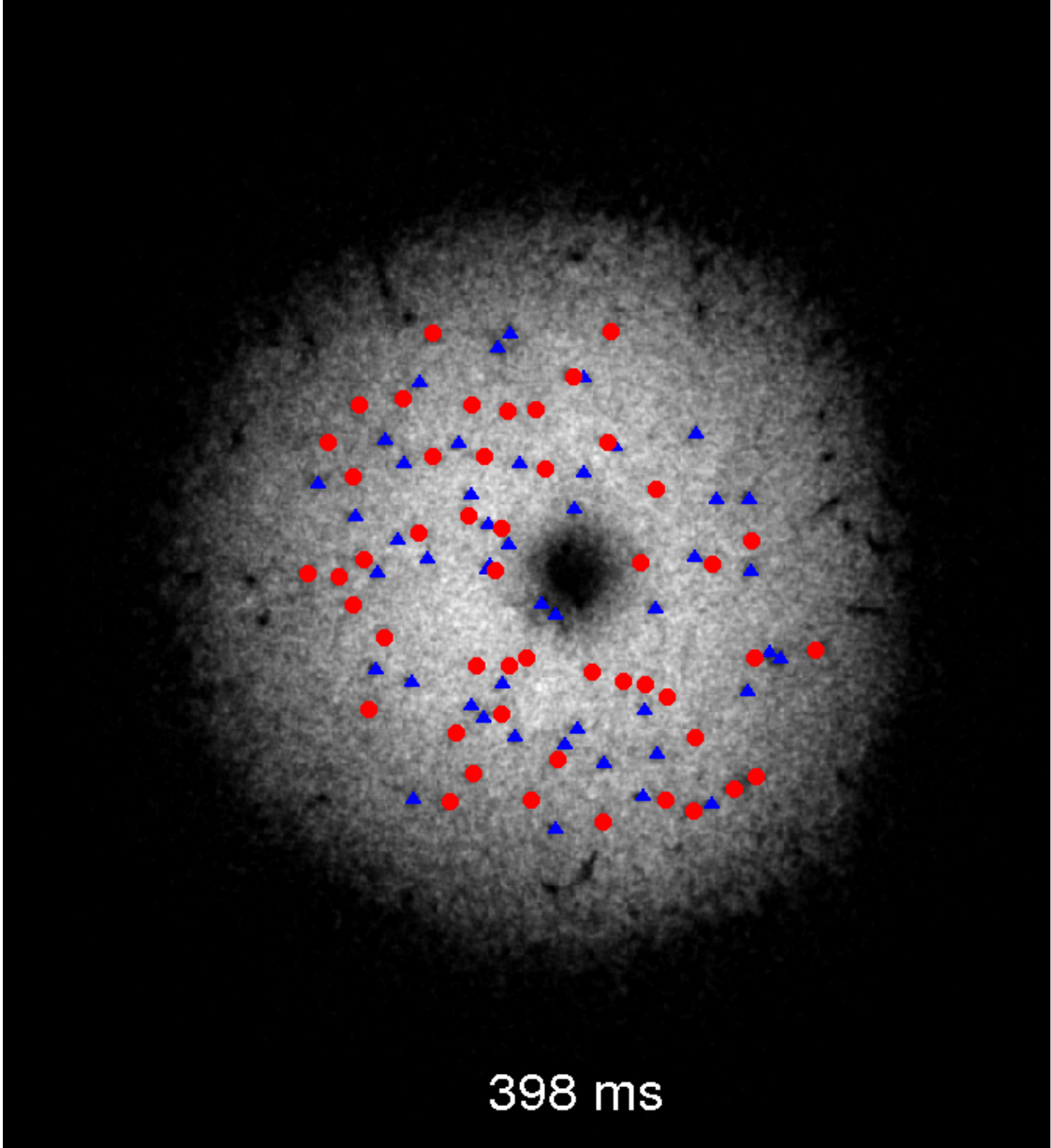}
	\hspace{-0.52in}\color{white}\raisebox{0.5ex}{\textsf{398~ms}}
	\includegraphics[width=0.22\linewidth,clip=true,trim=25 40 45 40]{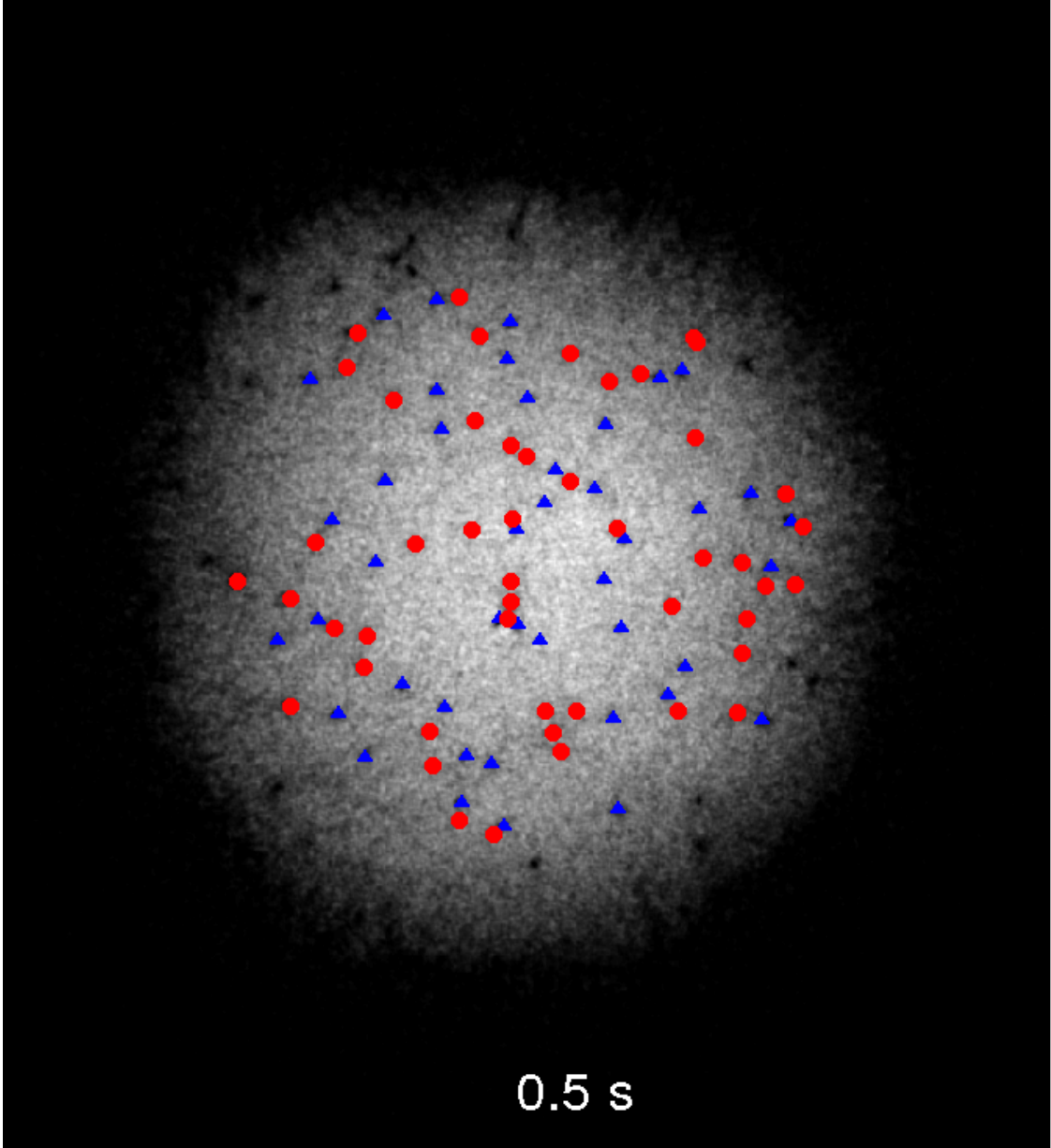}
	\hspace{-0.365in}\color{white}\raisebox{0.5ex}{\textsf{0.5~s}}
	\includegraphics[width=0.22\linewidth,clip=true,trim=20 40 50 40]{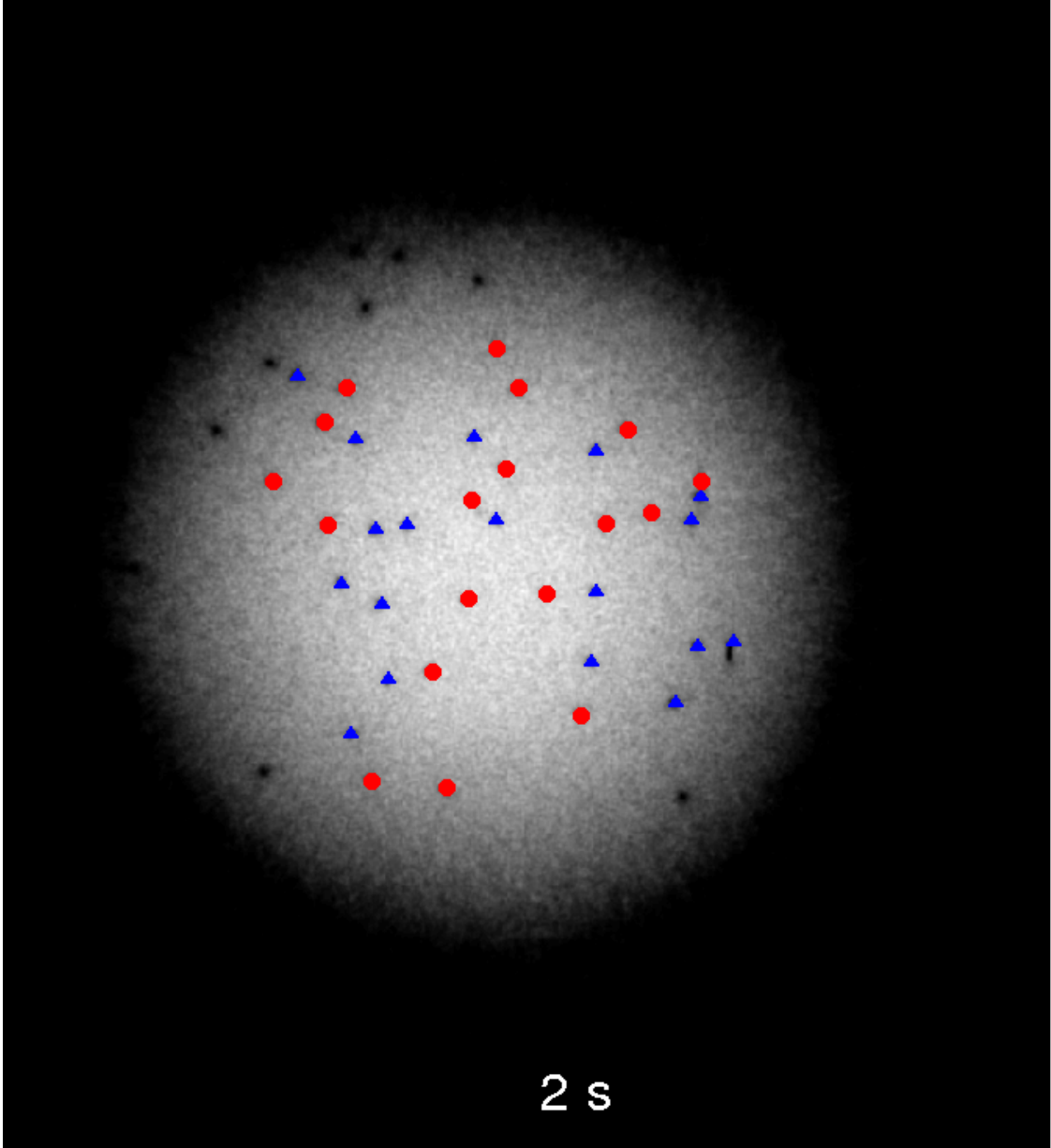}
	\hspace{-0.365in}\color{white}\raisebox{0.5ex}{\textsf{1.0~s}}
	\includegraphics[width=0.22\linewidth,clip=true,trim=20 40 50 40]{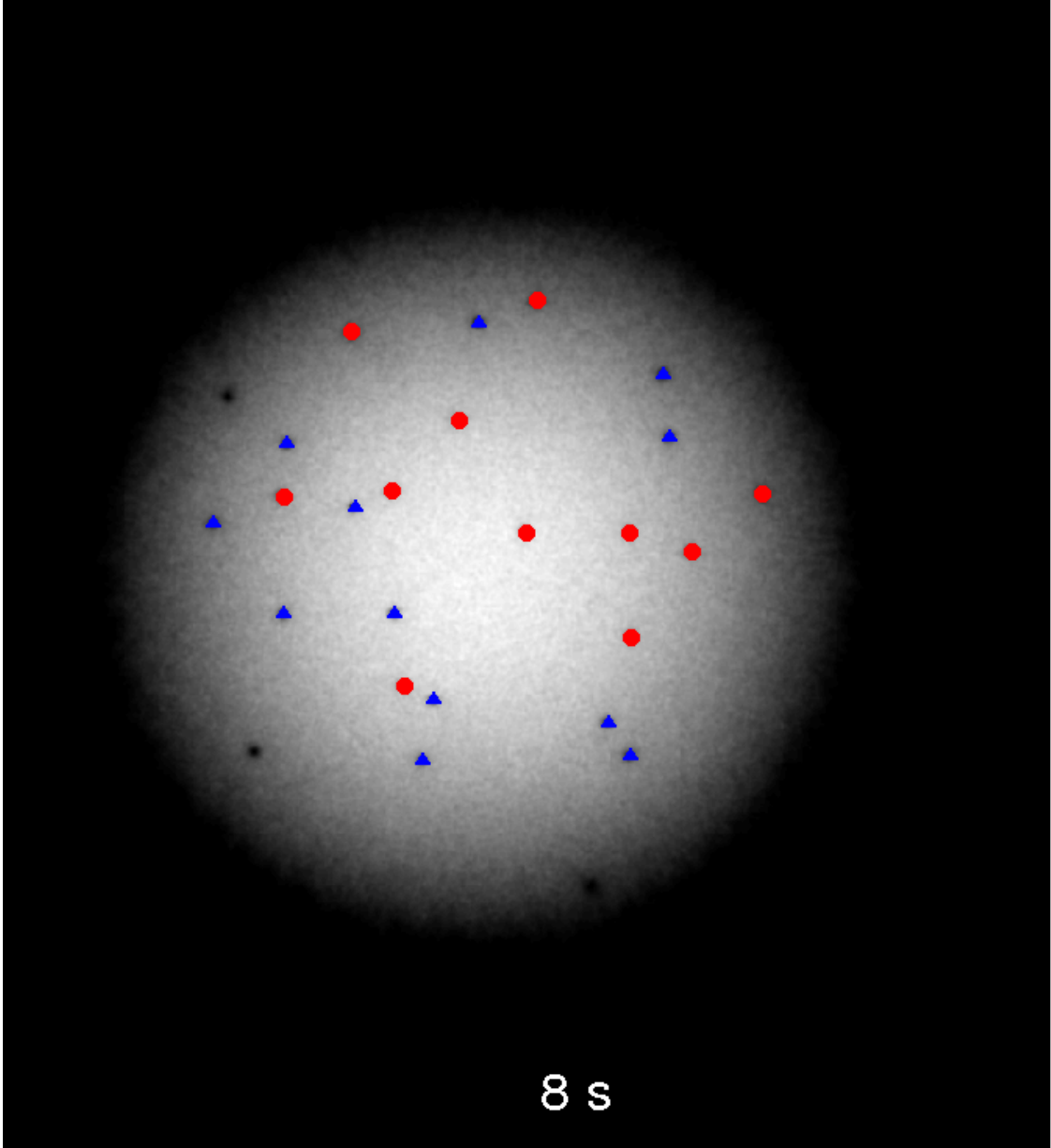}
	\hspace{-0.365in}\color{white}\raisebox{0.5ex}{\textsf{8.0~s}}\\
	\end{centering}
		\caption{Condensate flow past a circular obstacle with speed $v=1.4$mm/s, following the experiment of Ref. \cite{kwon_moon_14}. Each field of view is of size [$170\mu$m]$^2$, centered to best display the condensate.  Vortices with positive
(negative) circulation are highlighted by red circles (blue triangles).}\label{fig:circle08} 
\end{figure}
\begin{figure}
	\begin{centering}
	\includegraphics[width=0.22\linewidth,clip=true,trim=70 40 0 40]{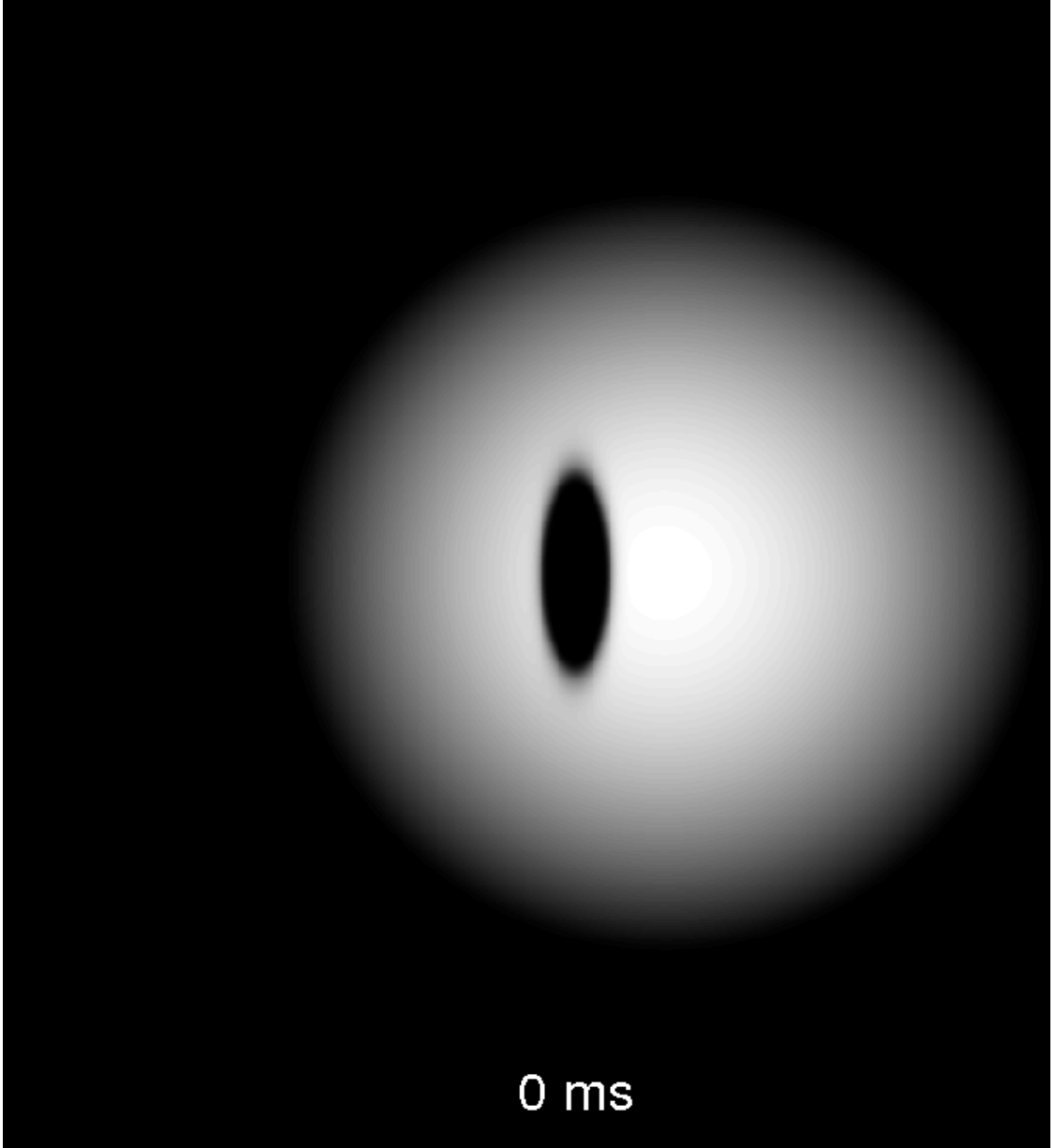}
	\hspace{-0.52in}\color{white}\raisebox{0.5ex}{\textsf{000~ms}}
	\includegraphics[width=0.22\linewidth,clip=true,trim=15 40 55 40]{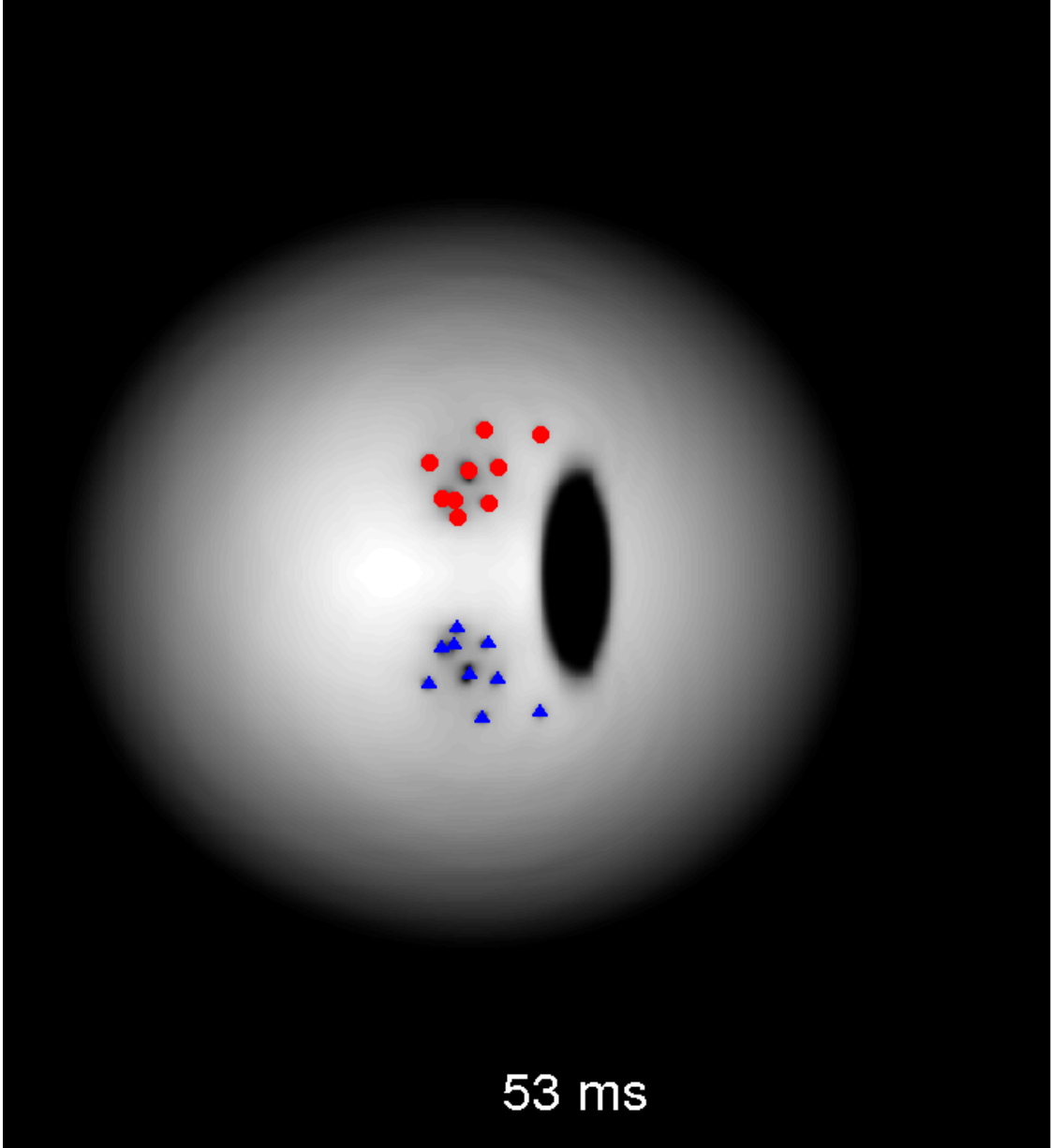}
	\hspace{-0.52in}\color{white}\raisebox{0.5ex}{\textsf{053~ms}}
	\includegraphics[width=0.22\linewidth,clip=true,trim=35 40 35 40]{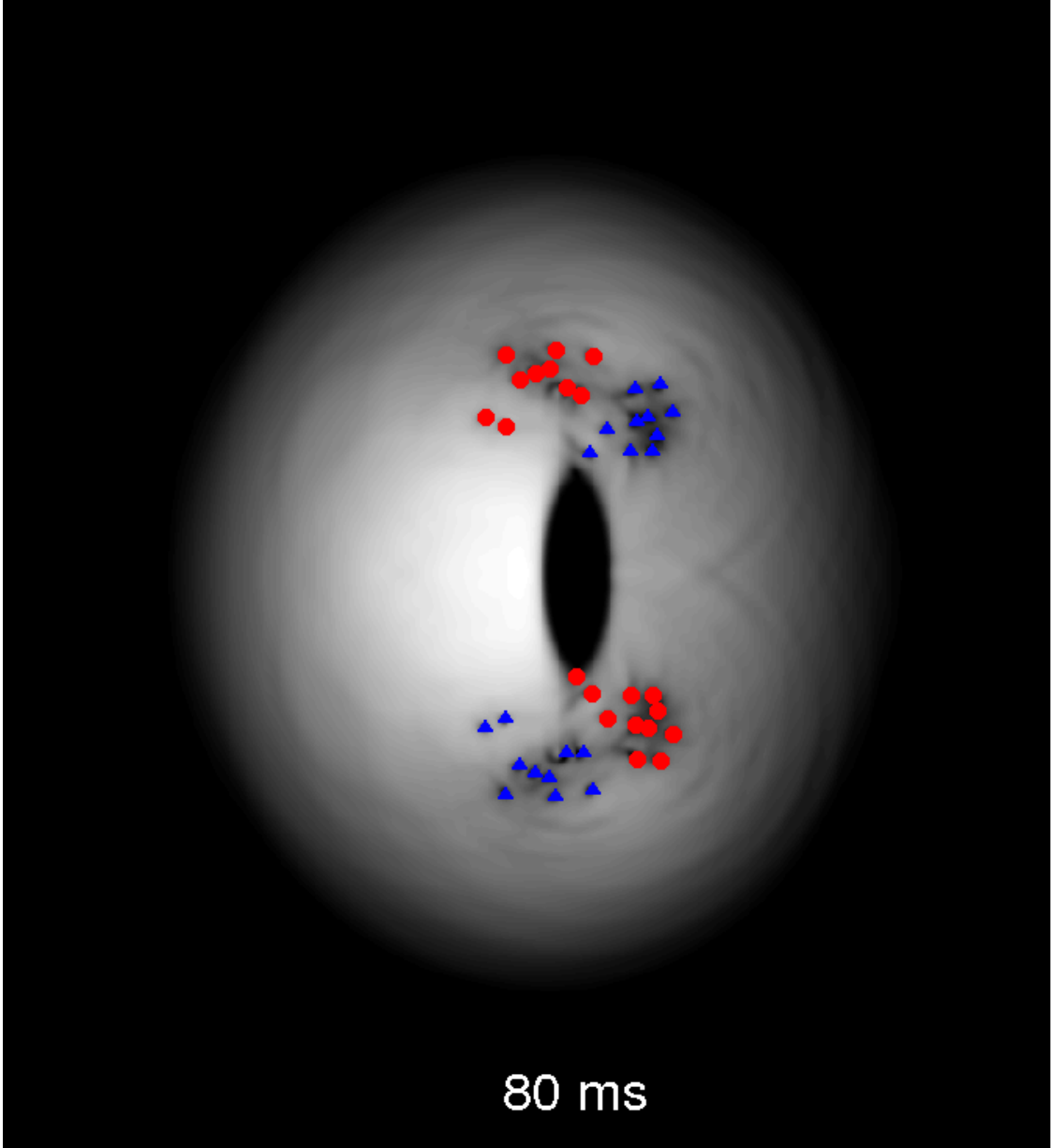}
	\hspace{-0.52in}\color{white}\raisebox{0.5ex}{\textsf{080~ms}}
	\includegraphics[width=0.22\linewidth,clip=true,trim=25 40 45 40]{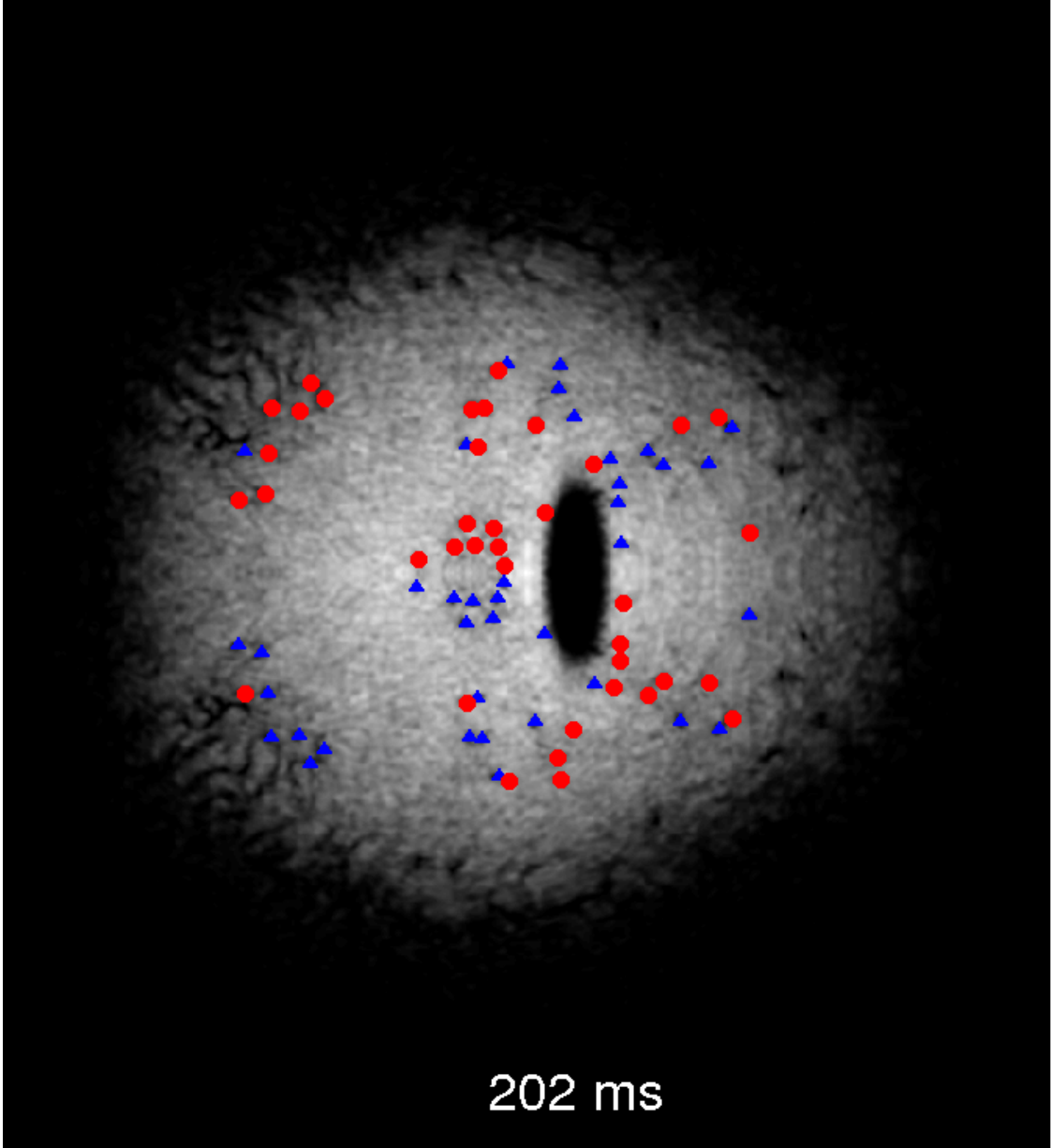}
	\hspace{-0.52in}\color{white}\raisebox{0.5ex}{\textsf{202~ms}}\\
	\includegraphics[width=0.22\linewidth,clip=true,trim=40 40 30 40]{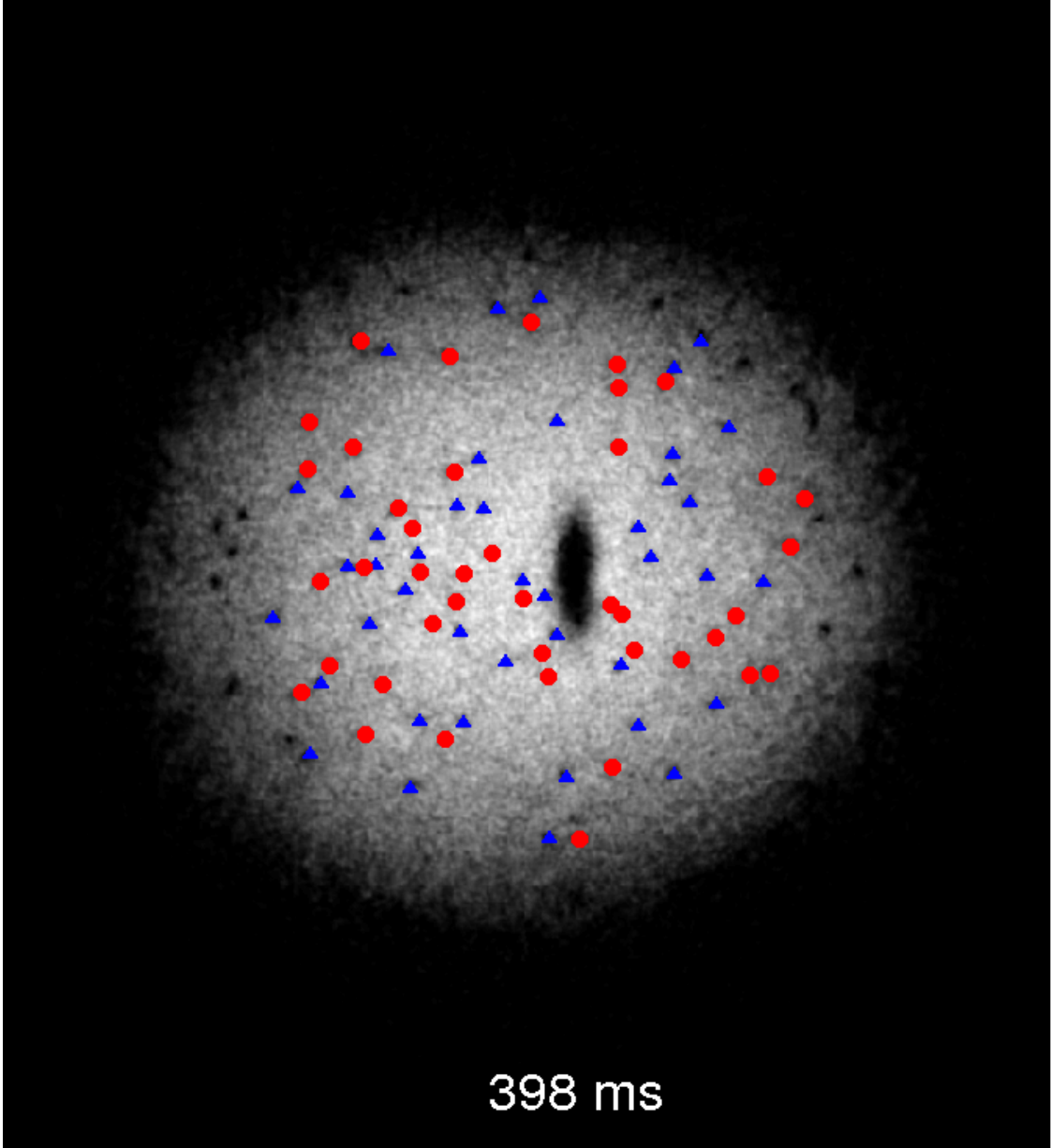}
	\hspace{-0.52in}\color{white}\raisebox{0.5ex}{\textsf{398~ms}}
	\includegraphics[width=0.22\linewidth,clip=true,trim=30 40 40 40]{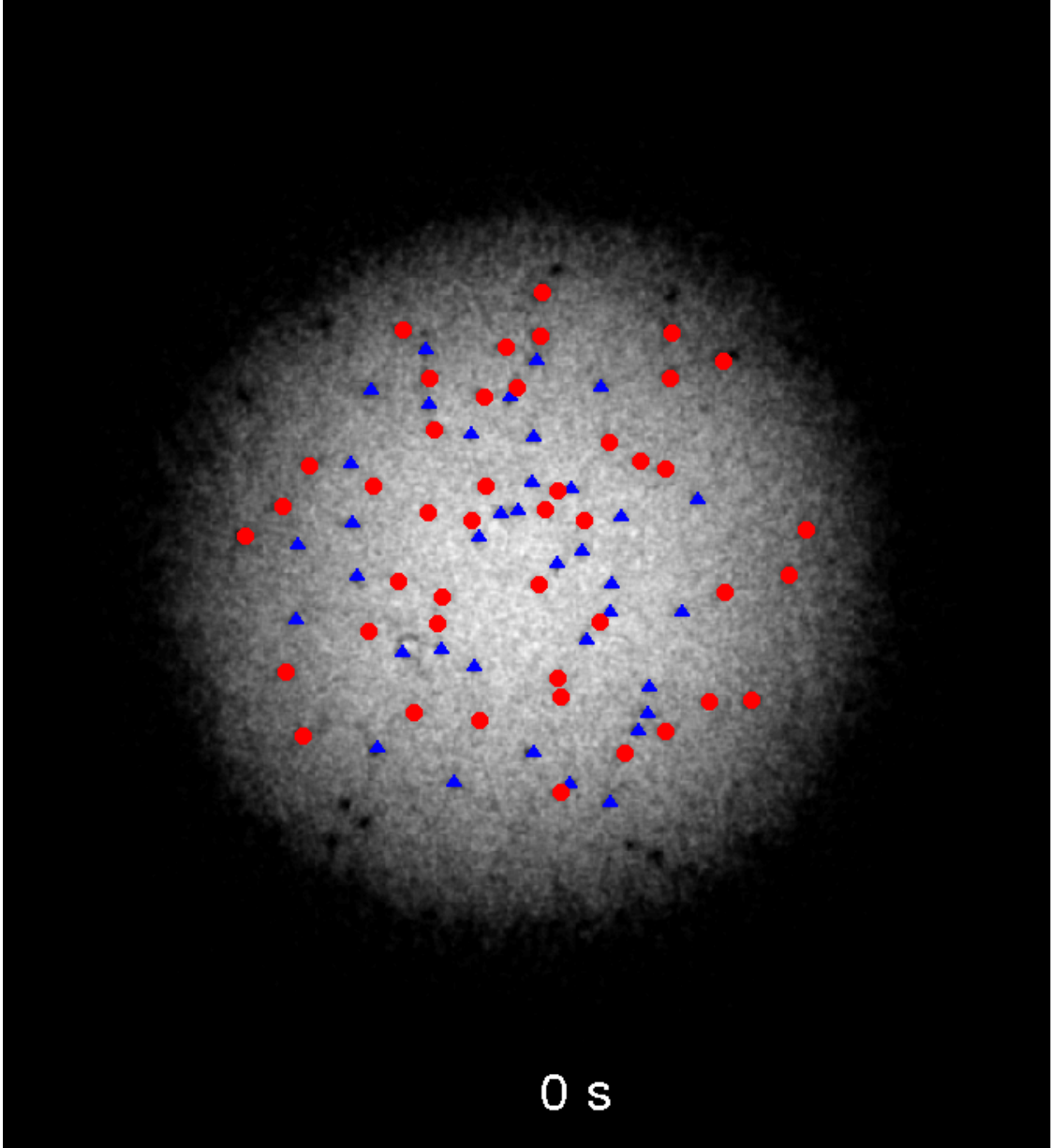}
	\hspace{-0.365in}\color{white}\raisebox{0.5ex}{\textsf{0.5~s}}
	\includegraphics[width=0.22\linewidth,clip=true,trim=25 40 45 40]{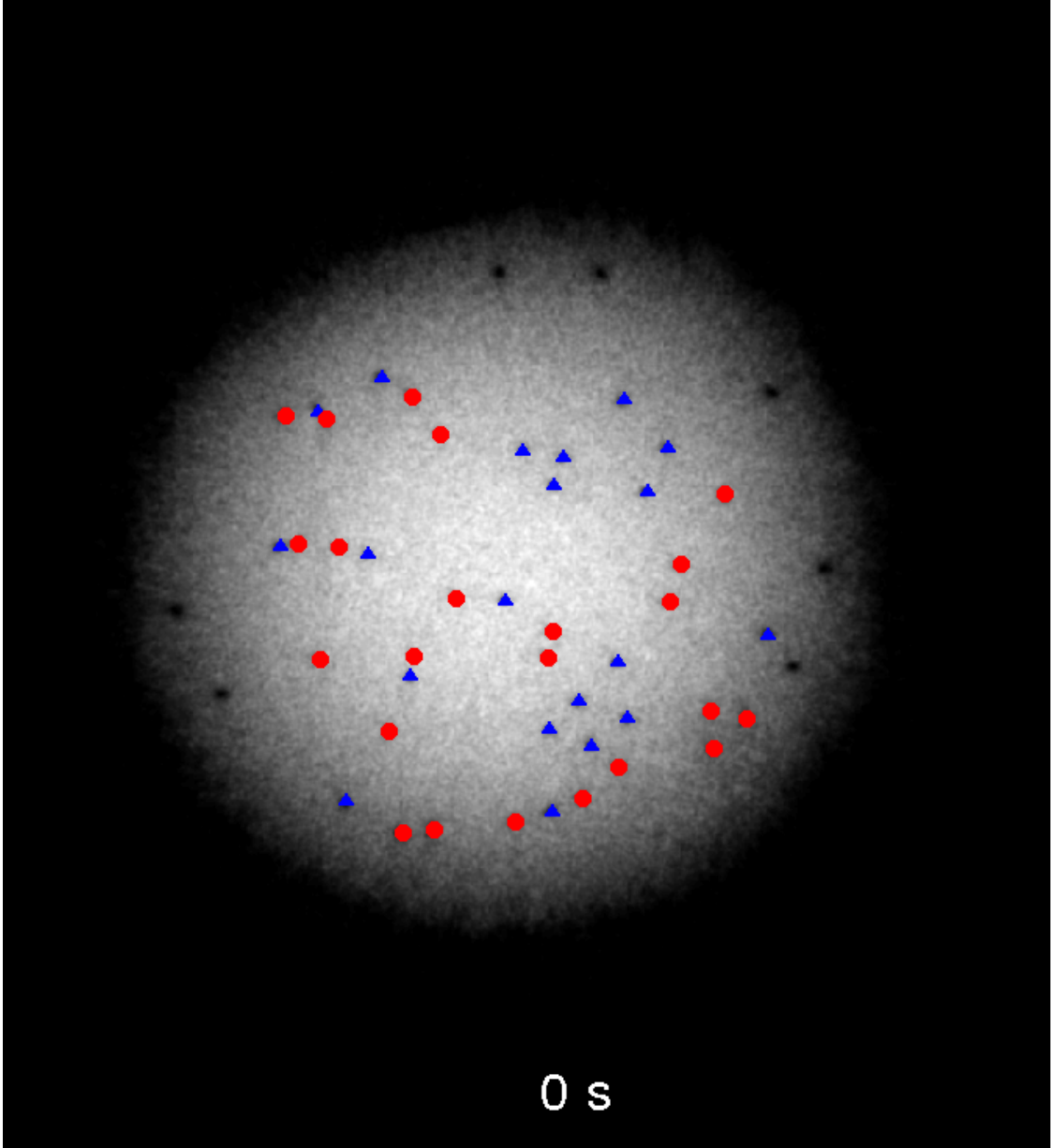}
	\hspace{-0.365in}\color{white}\raisebox{0.5ex}{\textsf{1.0~s}}
	\includegraphics[width=0.22\linewidth,clip=true,trim=25 40 45 40]{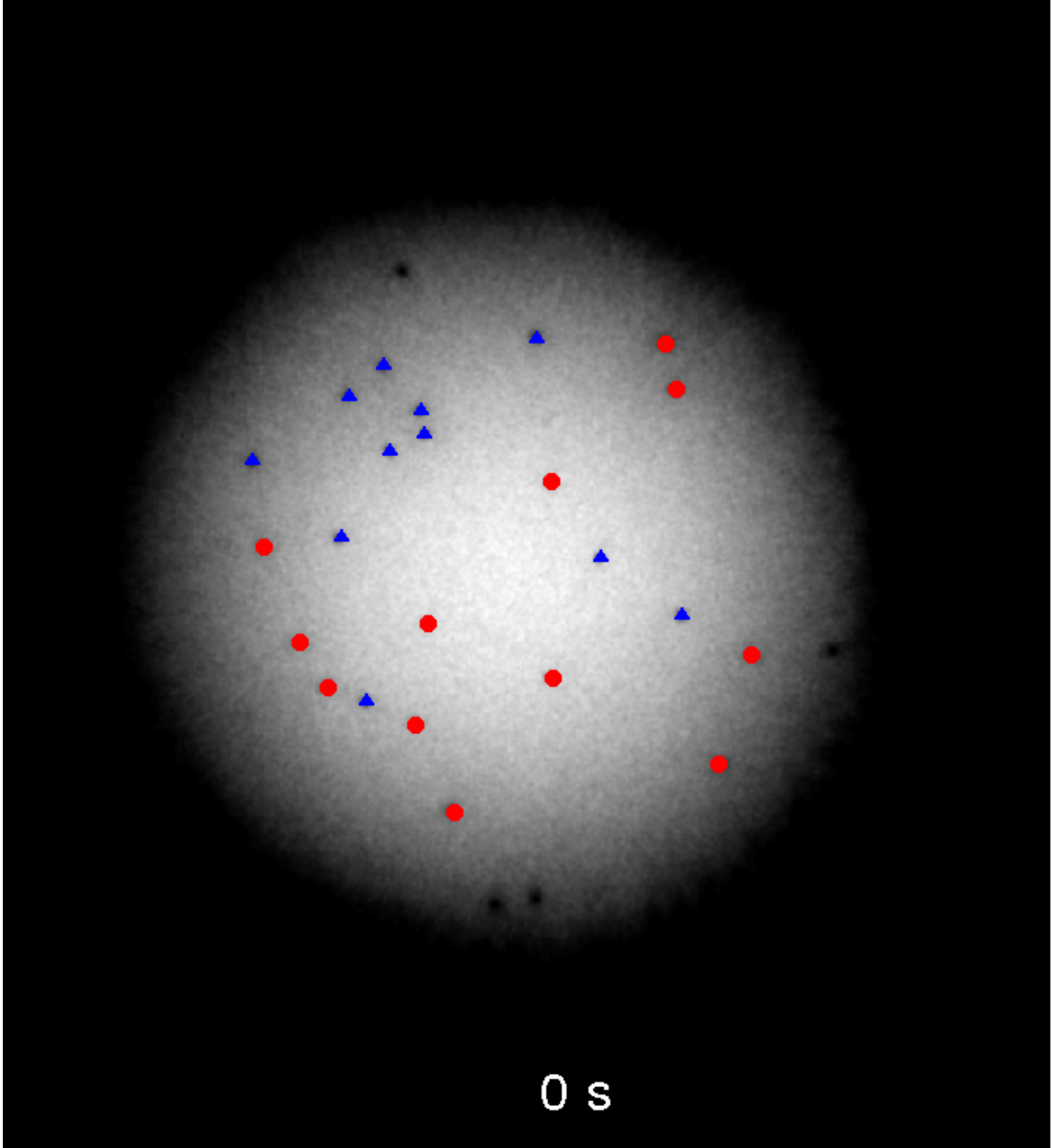}
	\hspace{-0.365in}\color{white}\raisebox{0.5ex}{\textsf{8.0~s}}\\
	\end{centering}
		\caption{\label{fig:circle08} As for Fig. \ref{fig:circle08} but with an elliptical obstacle $(\varepsilon=3)$ and a speed $v=0.8$mm/s.}
\end{figure}
}

Next we explore superfluid flow past an obstacle within a trapped condensate.  We follow the recent experiment by Kwon {\it et al.} \cite{kwon_moon_14}, which employed a highly-oblate and effectively 2D condensate with $1.8\times 10^6$ $^{23}$Na atoms. The in-plane harmonic trapping is circularly symmetric $V_{\rm{trap}}(x,y)=\frac{1}{2}m\omega_r^2 (x^2+y^2)$ with radial frequency $\omega_r=2\pi \times 15$ Hz.  A blue-detuned laser obstacle ($V_0=15\mu$ and $d=10.6\mu$m) pierces the condensate at the origin, following which the harmonic trap is shifted by $37 \mu$m in $x$ at speed $v$.  For reference, the speed of sound at the trap centre is $c_0=\sqrt{n_0 g / m}\approx 4.6$ mm/s (where the 0-subscript denotes peak values), decreasing further out in accord with the decreasing density as $c({\bf r})=c_0 \sqrt{n({\bf r})/n_0}$.  The laser obstacle is then adiabatically removed over 0.4s.  As observed experimentally \cite{kwon_moon_14}, and consistent with GPE simulations \cite{stagg_allen_14}, vortices become nucleated for $v \gappeq 0.5$mm/s, with the number of vortices generated increasing as $v$ is further increased.  

First we consider the circular obstacle employed in the experiment with $v=1.4$ mm/s, with the condensate evolution depicted in Fig. \ref{fig:circle08}.  The translation of the trap initiates a centre-of-mass mode of the condensate.  As the condensate first ``sloshes'' to the left past the obstacle, a large number of vortices are nucleated in an irregular manner from each pole (53~ms).  Significant energy is also driven into sound (density) waves, shock waves and shape excitations, which are visible in these plots.   Before the vortices are washed far downstream, the condensate sloshes to the right, generating further vortices and mixing together the old and new vortices (80~ms).  This process repeats several times as the condensate sloshes.  When the obstacle is removed a disordered arrangement of vortices is apparent (0.5~ms).  This is consistent with the experimental observations, and synonymous with a state of two-dimensional quantum turbulence.   Following this, there is a slow decay in the vortex number.  In our simulations, which are free of dissipation, this occurs through the annihilation of vortices of opposite circulation; experimentally \cite{kwon_moon_14}, and in the presence of dissipation terms to the GPE \cite{stagg_allen_14}, vortices will also undergo thermal dissipation whereby they drift to the condensate edge and are lost.

We now consider the case where the circular obstacle is replaced with an elliptical one (with $\varepsilon=3$) and (for reasons to be made evident shortly) a reduced speed $v=0.8$ mm/s.  The vortices are now nucleated in a {\it regular} fashion, forming two clusters of like-signed vortices (53~ms), much like in the homogeneous system.  No strong sound waves are visible in the condensate.  As the condensate changes direction, again two like-signed clusters are formed, but now with alternate polarity (80~ms).  Over time, these vortices mix together, and once the obstacle is removed, the condensate and vortex distribution is similar to the circular case. Indeed, the number of vortices in the condensate is essentially identical for both cases ($\sim 80$ vortices), despite the reduced speed for the ellipse. 

At the same speed $v$, the elliptical obstacle will typically generate a larger number of vortices than the circular obstacle.  For example, for $v=0.8$~mm/s, the ellipse and circle generate approximately 80 and 60 vortices, respectively (measured following removal of the obstacle).   However, for sufficiently high speeds ($v \gappeq 1$~mm/s) the enhancement due to the elliptical obstacle diminishes; in both cases the number of vortices driven into the condensate saturates.  This suggests that for a given condensate there exists a natural maximum vortex density which can be supported.  This limit is likely to be imposed by the quadratic increase of vortex annihilation events with vortex density \cite{kwon_moon_14}, which will act to cancel out the freshly-nucleated vortices.  

\section{Conclusions}
Through simulations of the GPE we have examined two-dimensional superfluid flow past an elliptical, rather than a circular, obstacle.  This simple and experimentally feasible modification enables control over the critical speed for vortex nucleation, the vortex nucleation rate, and the ensuing vortex interactions.  For an ellipse which is elongated perpendicular to the flow, the enhanced vortex interactions promote the formation of quantum wakes (composed of multiple singly-charged vortices) which closely resemble those of classical (viscous) flow past a cylinder.  Our observations \cite{stagg_parker_14}  suggest the emergence of classical behaviour from the presence of a sufficient number of quanta

When incorporated into the experimental scenario of Ref. \cite{kwon_moon_14}, such classical-like wakes are formed initially but become disrupted by the limited size and sloshing mode of the system.  Employing a condensate which is elongated in the direction of translation would promote the formation of longer-lived vortex clusters, and provide an experimental route to study these wakes and more general coherent structures of quantized vortices \cite{Laurie,reeves}.   Meanwhile, for the purposes of generating two-dimensional quantum turbulence, the elliptical obstacle allows for more efficient (more vortices for a given speed) and cleaner (reduced disruption due to sound and shock waves) nucleation of vortices.

\ack
This work made use of the facilities of N8 HPC provided and funded by the N8 consortium and EPSRC (Grant No. EP/K000225/1). The Centre is co-ordinated by the Universities of Leeds and Manchester.
\\


\begin{thebibliography}{10}

\bibitem{taneda41}
S. Taneda, J. Phys. Soc. Jpn. {\bf 11}, No. 3, 302 (1956).
\bibitem{taneda112}
S. Taneda, J. Phys. Soc. Jpn. {\bf 50}, No. 4, 1398 (1981).
\bibitem{nagib}
Photograph reproduced with permission by Thomas Corke and Hassan Nagib.
\bibitem{annett}  J. Annett, {\it Superconductivity, Superfluids and Condensates} (Oxford University Press, Oxford, 2004).
\bibitem{frisch92}
T. Frisch, Y. Pomeau, and S. Rica, Phys. Rev. Lett. {\bf 69}, 1644 (1992).

\bibitem{persistent} Ryu {\it et al}., Phys. Rev. Lett. {\bf 99}, 260401 (2007); S. Moulder {\it et al.}, Phys. Rev. A {\bf 86}, 013629 (2012).

\bibitem{jo}  R. Gati and M. K. Oberthaler, J. Phys. B {\bf 40}, R61 (2007). 

\bibitem{vortices} A. L. Fetter, J. Low Temp. Phys. {\bf 161}, 445 (2010).
\bibitem{Henn}  E. A. L. Henn {\it et al}., Phys. Rev. Lett. {\bf 103}, 045301 (2009).
\bibitem{Neely} T. W. Neely, \emph{et al.}, Phys. Rev. Lett. {\bf 104}, 160401 (2010).
\bibitem{kwon_moon_14} W.~J. Kwon {\it et al.}, arXiv:1403.4658 (2014).


\bibitem{qt} M. Tsubota, J. Phys.: Condens. Matter {\bf 21}, 164207 (2009); A. C. White, B. P. Anderson and V. S. Bagnato, Proc. Nat. Acad. Sci. USA {\bf 111}, 4683 (2014); A. J. Allen, N. G. Parker, N. P. Proukakis and C. F. Barenghi, J. Phys.: Conf. Ser. {\bf 544}, 012023 (2014).

\bibitem{Raman} C. Raman, \emph{et al.}, Phys. Rev. Lett. {\bf 83}, 2502 (1999).
\bibitem{Onofrio} R. Onofrio {\it et al.}, Phys. Rev. Lett. {\bf 85}, 2228 (2000).
\bibitem{Inouye} S. Inouye {\it et al.}, Phys. Rev. Lett. {\bf 87}, 080402 (2001).

\bibitem{stringari}L. P. Pitaevskii and S. Stringari, {\it Bose-Einstein Condensation} (Oxford University Press, Oxford, 2003).

\bibitem{saito10}
K. Sasaki, N. Suzuki, and H. Saito, Phys. Rev. Lett. {\bf 104}, 150404 (2010).

\bibitem{stagg_parker_14} G. W. Stagg, N. G. Parker and C. F. Barenghi, J. Phys. B {\bf 47}, 095304 (2014).

\bibitem{win01}
T. Winiecki, Ph.D. thesis, The University of Durham, 2001.

\bibitem{stagg_allen_14}  G. W. Stagg, A. J. Allen, N. G. Parker and C. F. Barenghi, arXiv:1408:3268 (2014).

\bibitem{Laurie}
A.W. Baggaley, J. Laurie, and C.F. Barenghi,
%{\em Vortex-density fluctuations, energy spectra, and vortical regions
%in superfluid turbulence},
Phys. Rev. Lett. {\bf 109}, 205304 (2012).

\bibitem{reeves} M. T. Reeves, T. P. Billam, B. P. Anderson and A. S. Bradley, Phys. Rev. A {\bf 89}, 053631 (2014).

\end{thebibliography}
\end{document}